\documentclass[mnsc,nonblindrev]{informs4} 

\OneAndAHalfSpacedXI 

\usepackage{endnotes}
\let\footnote=\endnote
\let\enotesize=\normalsize
\def\notesname{Endnotes}%
\def\makeenmark{$^{\theenmark}$}
\def\enoteformat{\rightskip0pt\leftskip0pt\parindent=1.75em
  \leavevmode\llap{\theenmark.\enskip}}

\usepackage[ruled,linesnumbered,noend]{algorithm2e}
\usepackage{booktabs}
\usepackage{multirow}
\usepackage{subfigure}
\usepackage{enumerate}
\usepackage{hyperref}
\hypersetup{
    colorlinks = true,
    linkcolor = cyan,
    citecolor = cyan,
    filecolor = cyan 
}
\usepackage{mathtools}
\newcommand\eqd{\stackrel{\text{d}}{=}}

\usepackage{natbib}
 \bibpunct[, ]{(}{)}{,}{a}{}{,}%
 \def\bibfont{\small}%

\usepackage{pgfplots}
\pgfplotsset{compat=1.18}
\usepackage{caption}
\TheoremsNumberedThrough     
\ECRepeatTheorems

\theoremstyle{TH}
\newtheorem{argument}{Argument}

\usepackage{dsfont}

\JOURNAL{Management Science}
\EquationsNumberedThrough    

\begin{document}


\RUNAUTHOR{Hong, Song, and Wang}

\RUNTITLE{Fast DES of Markovian Queueing Networks through Euler Approximation}

\TITLE{Fast Discrete-Event Simulation of Markovian Queueing Networks through Euler Approximation}

\ARTICLEAUTHORS{%
 \AUTHOR{L. Jeff Hong}
 \AFF{School of Management and School of Data Science, Fudan University, Shanghai 200433, China, \EMAIL{hong\_liu@fudan.edu.cn}
 } 
\AUTHOR{Yingda Song}
 \AFF{Antai College of Economics and Management, Shanghai Jiao Tong University, Shanghai 200030, China, \EMAIL{songyd@sjtu.edu.cn}
 }
\AUTHOR{Tan Wang}
\AFF{School of Management, University of Science and Technology of China, Hefei, Anhui 230026, China, \EMAIL{wang\_tan@ustc.edu.cn}}
} 

\ABSTRACT{
The efficient management of large-scale queueing networks is critical for a variety of sectors, including healthcare, logistics, and customer service, where system performance has profound implications for operational effectiveness and cost management. To address this key challenge, our paper introduces simulation techniques tailored for complex, large-scale Markovian queueing networks. We develop two simulation schemes based on Euler approximation, namely the backward and forward schemes. These schemes can accommodate time-varying dynamics and are optimized for efficient implementation using vectorization. Assuming a feedforward queueing network structure,  we establish that the two schemes provide stochastic upper and lower bounds for the system state, while the approximation error remains bounded over the simulation horizon.  With the recommended choice of time step, we show that our approximation schemes exhibit diminishing asymptotic relative error as the system scales up, while maintaining much lower computational complexity compared to traditional discrete-event simulation and achieving speedups up to tens of thousands times. This study highlights the substantial potential of Euler approximation in simulating large-scale discrete systems.
}
\KEYWORDS{discrete-event simulation; Euler approximation; Markovian queueing network}

\maketitle

\section{Introduction}

Many real-world systems are discrete in nature: they exhibit a finite or countably infinite number of states, and transitions between these states are triggered by a discrete sequence of events. Such discrete systems represent a fundamental class of dynamic systems found in various domains, from manufacturing and transportation to communication networks and healthcare. For instance, consider a hospital's operations. The system's state can be described by discrete variables, such as the length of the waiting line, availability of surgery rooms, or occupancy of hospital beds. The state transitions are triggered by events like patient arrivals, surgeries, and discharges. Discrete systems have wide applications, making them an important area of study in fields like control theory, operations research, computer science, and engineering. Understanding and analyzing these systems is vital for optimizing their performance and making informed decisions in complex, real-world scenarios. 

Within the realm of analyzing discrete systems, the discrete-event simulation (DES) method stands as a standard, and in many cases, the only approach. This simulation technique focuses on modeling and tracking the occurrence of events in the system, as well as the subsequent state transitions they induce. By meticulously recording the order and timing of events, DES provides a precise method for describing and assessing discrete systems. The event-driven methodology at the core of DES aligns seamlessly with the behavior of these systems, making it an indispensable tool for decision-making, process optimization, and resource allocation across a wide spectrum of industries and academic disciplines. For a comprehensive introduction to DES and its applications, we refer readers to \cite{bcnn:simulation}.

While DES is a versatile and widely employed tool, it is not without its limitations. One prominent issue arises when dealing with systems of large scale or complexity. In such cases, the large volume of events can overwhelm the simulation process, resulting in substantial computational costs. 
Moreover, because DES inherently operates in a sequential manner by adhering to the order and timing of events, it poses significant challenges when attempting to leveraging modern hardware capabilities like vectorization to accelerate the simulation process.
While efforts have been made to address these challenges,
the complex interactions of events often impose restrictions on the extent to which the acceleration techniques can be effectively employed. 
For an overview of research and challenges in this direction, we refer to \cite{fujimoto2016research}.

The rapid development of modern industry has given rise to plenty of largely scaled and complex systems that play important roles in society. The study of these systems holds both profound theoretical value and practical significance. Take a modern healthcare system for instance. On one hand, it comprises multiple layers of healthcare services, including primary care provided by family physicians, secondary care administered by specialists, and tertiary care offered at medical centers. On the other hand, each of these layers constitutes a complicated service network in itself. For example, the National University Hospital in Singapore, a tertiary medical center with over 1,200 beds, operates an inpatient department with 43 wards, 12 delivery rooms and 36 operating theatres, as well as an outpatient department with 40 specialist clinics and 18 service centres (\citealt{nuh}). The healthcare system serves a tremendous number of patients every day,  triggering a large volume of events, and hence it would be exceedingly time-consuming to run DES for simulating the system. 
Traditional DES is inadequate in effectively handling such systems, thereby motivating us to explore new methodologies.

To illuminate the exploration for fast simulation of discrete systems,  let's establish an analogy with the simulation of continuous systems. Continuous systems find extensive applications in various fields, including finance, biology, physics and engineering, where phenomena such as stock prices, neuronal activities, fluid flow and spread of quantities are often best represented by continuous-time continuous state processes, like diffusion processes. The simulation of continuous systems has been extensively studied in the literature. Take a one-dimensional diffusion process $\{X(t), 0\leq t\leq T\}$ as an example. Suppose $X$ is specified by the stochastic differential equation
    \[
    \setlength{\abovedisplayskip}{3pt}
\setlength{\belowdisplayskip}{3pt}
    dX\left( t \right) = a\left( {X(t),t} \right)dt + b\left( {X(t),t} \right)d{W_t},
    \]
where $W_t$ is the standard Brownian motion, $a$ and $b$ are bi-variate functions satisfying the common regularity conditions so that $X$ is well defined.  In a comparable spirit, we shall use the term ``event" in the context of continuous systems to signify an occurrence that changes the system's state $X$. As $X$ changes continuously, we confront two challenges in its simulation. Firstly, there is an uncountably infinite number of events between $\left[ {t,t + \Delta t} \right)$ for any $\Delta t>0$. Secondly, complex interactions among events become apparent; changes in $X(t)$ not only impact $X(t)$ itself but also set off subsequent changes in $X(t)$ through $a$ and $b$, giving rise to a complex sequence of interconnected modifications.

There are typically two approaches to simulate diffusion processes: exact simulation and Euler approximation. Exact simulation methods, which were pioneered by \cite{beskos2005exact}, aim to sample from the true probability distribution of targeted diffusion processes, and therefore, they have the advantage of generating unbiased estimators. 
However, these methods are often based on deep exploitation of the model structure and probabilistic properties, which invokes a substantial degree of mathematical complexity. This complexity often poses a barrier to their practical deployment, especially for large-scale problems involving multiple dimensions and correlated processes (\citealt{blanchet2020exact}). 
On the other hand, Euler approximation is a more popular approach to simulate diffusion processes in practice. 
It is within the spectrum of time-discretization methods, which discretize the simulation horizon into smaller time intervals and approximate the state changes within each interval. For a comprehensive introduction of the time-discretization methods, we refer readers to \cite{kloeden1992stochastic}. In comparison to the exact simulation methods, Euler approximation is easier to implement and has near-universal applicability.  It adeptly tackles the two challenges inherent in simulating $X(t)$ at the same time through two essential techniques:
\begin{itemize}
        \item \emph{aggregation of time interval}, which reduces the event counts between $\left[ {t,t + \Delta t} \right)$ by focusing on the cumulative effect of all events in the time interval, and
        \item \emph{decoupling of event interactions}, which removes interactions by fixing $X\left( t \right)$ so it remains constant in $\left[ {t,t + \Delta t} \right)$ and only changes at ${t + \Delta t}$.
\end{itemize}

If we draw an analogy between the simulation methods for discrete systems and continuous systems, it become evident that DES takes the role of a general approach for exact simulation of discrete systems. When there is only a moderate number of events and transitions over the simulation horizon,
DES can exactly simulate the state dynamics by taking the advantage of the discrete nature of the systems. However, the computational cost to do so increases substantially when the system becomes highly complex, mirroring the disadvantage encountered in exact simulation methods for continuous systems. 
In contrast, there has been limited discussion on the time-discretization approximation methods for simulating discrete systems, possibly due to concerns about the induced error. Time discretization inherently results in a loss of information, as it preserves only the aggregate-level information in each time interval. Yet, as the system scales up, the importance of microscopic-level information diminishes. Managers often prioritize aggregate-level information, such as summary statistics, for comprehending the overall system performance. This is precisely where the strength of Euler approximation becomes evident. 
It is therefore interesting to see whether we can employ the techniques of Euler approximation to overcome the challenges in simulating large-scale and complex discrete systems. Specifically, we hope to leverage the aggregation of time intervals to reduce the total computational effort, and to apply the decoupling of event interactions to facilitate vectorization, which allows us to further take advantage of efficient vector computational tools.

In this paper, we explore the possibility of simulating discrete systems through the lens of Euler approximation, with a specific focus on Markovian queueing networks. This choice is motivated by the following reasons.  Firstly, the memoryless property of the Markovian queueing networks simplifies the design and analysis of the corresponding simulation schemes, making them widely accepted benchmark models to study discrete system simulations. For example, see \cite{shahabuddin1994importance}, \cite{andradottir2003efficiency} and \cite{Busic2015}.
Secondly, under specific assumptions, Markovian queueing networks possess the product-form property, and we have analytical solutions for their steady-state performance measures. These solutions can be used to validate the effectiveness of our approach. Lastly, our method extends to scenarios where analytical solutions are unavailable, for example, when the input parameters are time-varying. In such instances, our approach stands as an effective tool for evaluating the system performance. 

There is a substantial body of literature on analytical and approximate methods for solving Markovian queueing networks. While some methods can be highly efficient for particular types of problems, their practical use is often limited to cases that make certain simplified assumptions about the network's structure, the kinds of performance measures to be evaluated, the system scale, or are confined to systems that operate under stationary conditions, yielding only steady-state analyses. When it comes to solve models that exhibits high fidelity with real-world scenarios, simulation analysis may become a necessity (\citealt{glynn2022queueing}). This paper represents an initial exploration into the use of Euler approximation for simulating large-scale discrete systems. While currently focused on Markovian systems, we are optimistic about the potential of our approach, and intent to refine and expand our methods to general non-Markovian systems in the future.

We emphasize that the development of Euler approximation for Markovian queueing network is a non-trivial task. It gives rise to two central research questions that require thorough investigation, addressing which is essential not only in the context of Markovian queueing networks but also for discrete systems in general. 
(1) \emph{How to effectively perform Euler approximation for discrete systems like queueing networks?} Euler approximation is originally designed for time-driven continuous systems, and thus it is not clear how to adapt this method for simulating event-driven discrete systems. Queueing networks involve events such as arrivals, departures, and service completions. Since multiple events can occur within a given time interval, determining the aggregate impact of these events on state transitions requires careful consideration. Furthermore, for queueing networks with a high level of complexity, featuring multiple layers,  numerous servers, and time-varying dynamics, the interplay of events over multiple queues can become intricately intertwined. Such coordination complexity presents further challenges for the successful application of vectorization techniques.  
(2) \emph{What's the approximation error incurred by time discretization and how to manage it?}
Time discretization inevitably introduces approximation error. Intuitively, the degree of discretization error hinges on the choice of the time step. A very small time step may lead to excessive computational costs, while a large time step may lead to inaccurate simulations. Moreover, when the system scales up, so do the errors.
Therefore, it is crucial to select an appropriate time step which could strike a balance between efficiency and accuracy, particularly for large-scale queueing systems.

Our paper contributes mainly on three aspects. Firstly, we develop two simulation schemes for Markovian queueing networks based on Euler approximation: the backward scheme and the forward scheme. These are designed to accommodate time-varying dynamics and are optimized for efficient implementation using vectorization techniques.
Secondly, under the assumption of feedforward network structure, we prove that these simulation schemes yield stochastic upper and lower bounds for the state of the original queueing network and establish the theoretical error bounds. It is a pleasant surprise to find that the magnitude of the approximation error does not propagate across the simulation horizon, which makes it different from the Euler approximation of diffusion processes.
Thirdly, for large-scale Markovian queueing networks, we establish a guideline for selecting an appropriate time step based on the asymptotic analysis of relative errors. Our results indicate that the recommended time step ensures the asymptotic relative error of our simulation schemes approaches zero as the network scales up,  while maintaining a much lower computational complexity compared to the traditional DES.

The rest of the paper is organized as follows. Section~\ref{sec:sim_state} presents our approach for simulating Markovian queueing networks with Euler approximation. It starts with a single-station queue and then extends to general Markovian queueing networks. Section~\ref{sec:pure_departure} is dedicated to the discussion of a key component of our approach -- the pure departure processes. Section~\ref{sec:error} studies the approximation error and derives the error bound. It further analyzes the asymptotic behavior of the relative approximation errors and discusses the selection rule for time step.
Section~\ref{sec:sojourn} extends our approach to the simulation of customer sojourn time. The numerical results are provided in Section~\ref{sec:exp}. All proofs are deferred to the online appendices.

\section{Simulating the State Dynamics of Markovian Queueing Networks with Euler Approximation}\label{sec:sim_state}
In a Markovian queueing network, the system state refers to the numbers of customers in each queue. 
Similar to the traditional Euler approximation to simulate diffusion processes, our approach starts by selecting a time step $h>0$ (the selection criteria for $h$ is discussed in Section \ref{subsec:asymptotic_error}), and updates the system state at time $t=\tau h$ $(\tau=1,2,\cdots)$ progressively. For clarity, we first present our approach for simulation of a single-station queue in Section~\ref{subsec:sim_single_station}, and then extend it to general multi-layer queueing networks in Section~\ref{subsec:sim_net}.

\subsection{Simulation of Single-Station Queue}\label{subsec:sim_single_station}

Consider a single service station modelled as a multi-server queue. Denote $\lambda$ as the arrival rate, $\mu$ as the service rate, and $m$ as the number of servers. In our discussion, we assume $\mu$ is constant, but we allow $\lambda$ and $m$ to be \emph{time-varying}. This is motivated by the fact that many practical scenarios exhibit dynamic customer demand, which can result in time-varying arrival rates. Furthermore, staffing servers according to the time-varying demand is also a common feature in real systems to achieve operational efficiency. For instance, consider call centers where the volume of incoming calls fluctuates over time with a repetitive pattern. In such case, it's a standard practice to utilize pre-designed staffing schemes for server allocation within specific time horizons, in order to maintain a consistent level of service efficiency. Our chosen setting aligns closely with these real-world characteristics.

Given the number of customers in the station is $N_{\tau-1}$ at time $t=(\tau-1)h$, we aim to simulate the number of customers $N_\tau$ in the same station at time $t=\tau h$. It is possible to perform exact simulation by DES or deriving the probability mass function of $N_\tau$ by solving ordinary differential equations, but we are interested to develop a simple approximation scheme that can be implemented efficiently even for large-scale problems. In general,
$N_{\tau}$ is updated from $N_{\tau-1}$ through the equation
$$
\setlength{\abovedisplayskip}{3pt}
\setlength{\belowdisplayskip}{3pt}
N_\tau = N_{\tau-1}+A_\tau-D_\tau,
$$
where $A_\tau$ and $D_\tau$ represent the aggregated number of arrivals and departures from time $(\tau-1)h$ to time $\tau h$, respectively.
Therefore, it suffices to sample $A_\tau$ and $D_\tau$.

\subsubsection{Sampling of $A_\tau$}\label{subsubsec:sim_A}
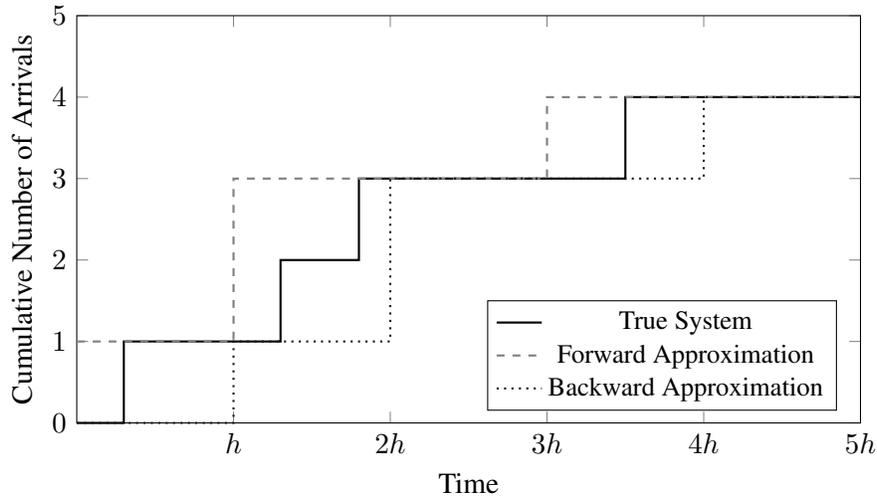
\begin{figure}[htb]
\begin{center}

\begin{tikzpicture}
\begin{axis}[
    xlabel={Time},
    ylabel={Cumulative Number of Arrivals},
    xmin=0, xmax=5,
    ymin=0, ymax=5,
    ytick={0,1,2,3,4,5},
    xtick={1,2,3,4,5}, 
    xticklabels={$h$, $2h$, $3h$, $4h$, $5h$}, 
    grid style=dashed,
    legend pos=south east, 
    legend style={font=\small},
    width=12cm, 
    height=7cm,
    clip=false
]

\addplot+[const plot, no marks, thick, black] coordinates {
  (0,0) (0.3,1) (1.3,2) (1.8,3) (3.5,4) (5,4)
};
\addlegendentry{True System}

\addplot+[const plot, no marks, thick, gray, dashed] coordinates {
    (0,1) (1,3) (3,4) (5,4)
};
\addlegendentry{Forward Approximation}

\addplot+[const plot, no marks, thick, black!90, dotted] coordinates {
   (0,0) (1,1) (2,3) (4,4) (5,4)
};
\addlegendentry{Backward Approximation}
\end{axis}
\end{tikzpicture}

\caption{A Sample Path of Arrival Processes for the True System and Two Euler Approximation Schemes.} \label{fig:sampling_A}
\end{center}
\end{figure}

Under the Markovian assumption, simulating the aggregated number of arrivals within a given time interval is straightforward. If the arrival rate at time $t$ is $\lambda(t)$, then $A_\tau$ follows a Poisson distribution with expected value of $\int_{(\tau-1)h}^{\tau h}\lambda(t)dt$. In this way, the sequence $\{A_\tau, \tau\geq 1\}$ is sampled exactly from the joint distribution of the aggregated numbers of arrivals, with approximation errors stemming from the loss of precise arrival times for each individual customer. To address this, we introduce two Euler approximation schemes: \emph{the backward approximation} and \emph{the forward approximation}, representing two opposite extreme scenarios. The backward approximation assumes all the arrivals within the interval $((\tau-1)h, \tau h]$ occur simultaneously at $\tau h$. In essence, it shifts all arrival events backward to the end of the time interval. On the contrary, the forward approximation assume the opposite, where all arrivals are considered to happen at $(\tau-1)h$, effectively advancing all arrival events forward to the start of the interval.
As a result, the cumulative count of arrivals to the system is consistently capped by the forward approximation and bounded from below by the backward approximation. However, at the end of each time interval, the cumulative counts of arrivals for the actual system and the two approximation schemes align exactly (see Figure~\ref{fig:sampling_A} for an illustration).

It is important to note that our methodology is adaptable and can be expanded to accommodate more complex arrival processes. Firstly, the Markovian assumption on the arrival process is not a necessity in our approach, as long as we can sample the aggregated number of arrivals within each time interval. For example, when the arrival process is a doubly stochastic Poisson process \citep{oreshkin2016rate,zheng2023doubly}, we can simulate $\{A_\tau, \tau\geq 1\}$ by conditioning on the random arrival intensity.
Secondly, our time aggregation technique for arrivals is particularly compatible with situations where simulation models are driven by empirical data. Take, for example, the modeling of an arrival process for an online service center using historical data that records the number of arrivals for each minute within the operating hours over the previous week. For such cases, our approach would involve utilizing established methods to draw samples of aggregated arrivals for forthcoming minutes from the empirical distribution. This strategy is evidently more straightforward and robust than estimating the arrival rate function and implementing conventional simulation techniques like DES.

\subsubsection{Sampling of $D_\tau$}\label{subsubsec:sim_D}

Given that the number of arrivals within the time interval is  $A_{\tau}$, next we simulate the number of departures $D_\tau$ within the same time interval. As outlined in Section~\ref{subsubsec:sim_A}, our approximation schemes assume all the arrival events occur at either the beginning or the end of each interval.  As a result, the evolution of the system state within any given time interval involves only departure events, and hence we call it a \emph{pure departure process}. Therefore, our essential task is to simulate how many customers could finish service for this pure departure process. It should be noted that this task is far from straightforward,  as the service rate of the station depends on the number of busy servers, which may fluctuate randomly across the interval. 

In Section~\ref{sec:pure_departure_simu} 
we propose an exact simulation scheme (Algorithm~\ref{alg:pure-departure}) to sample departure from a pure departure process.  It takes several parameters as input: the initial system state $x$, the number of servers $m$, the service rate $\mu$, and the length of the time interval $h$. It then produces a random sample on the number of departures as the output. As will be further explained in Lemma~\ref{lem:pure-departure-1} in Section~\ref{sec:pure_departure}, this algorithm is exact for simulating the pure departure process. In the remainder of this paper, we will use the following notation to denote the generation of a departure sample using this algorithm:
\[
\setlength{\abovedisplayskip}{3pt}
\setlength{\belowdisplayskip}{3pt}
D \sim \text{GenerateDeparture}(x,m,\mu,h).
\]

Note that the algorithm is readily adaptable to scenarios in which the number of servers is time-varying over the targeted simulation horizon. For example, consider the task of simulating the number of customers at a single station over a time horizon of $t$, with server allocation being adjusted at intermediate time points $\{t_1, \cdots, t_k\}$. 
Then at the initial stage of our simulation procedure, we can discretize the time horizon such that these time points are the boundary points of our time intervals. This ensures a constant number of servers within each time interval, allowing for exact simulation of departures using Algorithm~\ref{alg:pure-departure}. 

A particular challenge arises when managing time-varying numbers of servers, especially if a server is scheduled to be withdrawn from service while actively serving a customer. In such scenarios, our current algorithm reassigns the interrupted customer to the waiting queue by default. This approach is a simplification designed to facilitate simulation, but it may not accurately reflect more complex operational strategies without modification. For instance, a customer who experiences a service interruption might be granted higher priority in subsequent service periods.
This challenge is rooted in the study of queueing systems that involve service interruptions and preemptive priority disciplines, which trace back to the seminal paper by \cite{white1958queuing}. We recommend textbooks on queueing theory such as \cite{bhat2015introduction} to gain a comprehensive understanding of these issues. Our simulation algorithm is flexible and can be tailored to match specific operational policies, thus ensuring fidelity between the simulation and actual system dynamics.

\subsubsection{Sampling of $N_\tau$}\label{subsubsec:sim_N}

Combine the procedures developed in the previous two subsections, we are able to provide two approximation schemes for simulating the system state $N_\tau$. 

\emph{The backward Euler approximation scheme} aggregates arrivals backward to the end of each time interval, therefore the number of departures is sampled from a pure departure process with initial value $N_{\tau-1}^b$ before the arrivals occur.
Starting from the initial state $N_0^b = N_0$, the scheme updates the system state according to
$$
\setlength{\abovedisplayskip}{3pt}
\setlength{\belowdisplayskip}{3pt}
N_\tau^b = N_{\tau-1}^b+A_\tau-D_\tau^b, \quad \text{where } D_\tau^b\sim \mathrm{GenerateDeparture}(N_{\tau-1}^b, m, \mu, h),
$$
for $\tau=1,2,\cdots.$

\emph{The forward Euler approximation scheme} aggregates arrivals forward to the start of each time interval. Thus, the number of departures is sampled from a pure departure process with initial value $N_{\tau-1}^f+A_\tau$ after $A_\tau$ is realized. Starting from the initial state $N_0^f = N_0$, the scheme updates the system state according to
$$
\setlength{\abovedisplayskip}{3pt}
\setlength{\belowdisplayskip}{3pt}
N_\tau^f = N_{\tau-1}^f+A_\tau-D_\tau^f, \quad \text{where } D_\tau^f\sim \mathrm{GenerateDeparture}(N_{\tau-1}^f+A_\tau, m, \mu, h),$$
for $\tau=1,2,\cdots.$

\subsection{Simulation of Queueing Networks}\label{subsec:sim_net}

Consider a general queueing network with $n$ nodes, each representing a service station modeled as a first-in-first-out multi-server queue. After completing service at the $i$th node, a customer either transitions to the $j$th node with probability $p_{ij}$ ($j=1,\cdots,n$) or exits the system, represented by the state $\partial$, with probability $p_{i\partial}$. The probabilities are such that $\sum_{j=1}^{n} p_{ij} + p_{i\partial} = 1$.
Furthermore, the network is assumed to have Markovian properties, with customers arriving at the $i$th node from outside the system following a Poisson process with rate $\lambda_i$. Each of the $m_i$ servers at the $i$th node has an exponential service rate $\mu_i$. As in our previous discussion of a single-station queue, we assume that $\mu_i$ is constant over time for any fixed $i$, while $\lambda_i$ and $m_i$ may vary with time. The number of servers $m_i$ remains unchanged within each time interval for Euler approximations.

The primary difference between simulating a node within a queueing network and a single-station queue is the source of arrivals. In a single-station queue, all customers arrive from outside the system, while in a queueing network, a node receives customers both externally from outside the system and internally from other nodes in the network. To differentiate these arrivals, we use notations $A_{\tau,i}^{ex}$ and $A_{\tau,i}^{in}$ to represent the number of external and internal arrivals, respectively, within the $\tau$th time interval, where the subscript $i$ represents the index of the node in concern.

The process for sampling the number of external arrivals $A_{\tau,i}^{ex}$ to node $i$ is analogous to the method detailed in Section~\ref{subsubsec:sim_A}. For internal arrivals to node $i$, the quantity $A_{\tau,i}^{in}$ is determined by the equation:
\[
\setlength{\abovedisplayskip}{5pt}
\setlength{\belowdisplayskip}{5pt}
A_{\tau,i}^{in} = \sum_{k=1}^{n} {{R_{\tau,ki}}},
\]
where $R_{\tau,ki}$ denotes the number of customers transitioning from node $k$ to node $i$ during the $\tau$th time interval. 
Furthermore, the conditional joint distribution of $(R_{\tau,k1}, \cdots, R_{\tau,kn}, R_{\tau,k\partial})$ given the number of total departures from node $k$, i.e., $D_{\tau,k}$, follows a multinomial distribution
\[
\setlength{\abovedisplayskip}{3pt}
\setlength{\belowdisplayskip}{3pt}
(R_{\tau,k1}, \cdots, R_{\tau,kn},R_{\tau,k\partial},)| D_{\tau,k} \sim \mathrm{Multinomial}(D_{\tau,k}, (p_{k1},\cdots, p_{kn},p_{k\partial})).
\]
Therefore, the sampling of internal arrivals to node $i$ during the $\tau$th time interval hinges on the departures from its upstream nodes within the same interval.

By assuming all the external and internal arrival events occur at the end of each time interval, we establish the backward Euler approximation schemes for simulating queueing networks. During the $\tau$th time step, the sequence of tasks to be performed at the $i$th node are:
\begin{enumerate}
\item generate the number of departures $D^b_{\tau,i}\sim\mathrm{GenerateDepartures}(N^b_{\tau-1,i},m_{\tau,i},\mu_i,h)$;
\item generate $(R^b_{\tau,k1}, \cdots, R^b_{\tau,kn},R^b_{\tau,k\partial},)| D^b_{\tau,k} \sim \mathrm{Multinomial}(D^b_{\tau,k}, (p_{k1},\cdots, p_{kn},p_{k\partial}))$;
\item generate the number of external arrivals  $A^{ex}_{\tau,i}\sim \mathrm{Poisson}\left(\int_{(\tau-1)h}^{\tau h}\lambda_i(u)du\right)$, and compute the number of internal arrivals $A_{\tau,i}^{in,b} = \sum_{k=1}^{n} {{R^b_{\tau,ki}}}$;
\item update the state $N^b_{\tau,i}= N^b_{\tau-1,i}-D^b_{\tau,i}+A^{ex}_{\tau,i}+A^{in,b}_{\tau,i}$.
\end{enumerate}
It's crucial to recognize that calculating internal arrivals at node $i$ during Step 3 depends on the routing outcomes from Step 2 at other nodes indexed by $k$ with $p_{ki}>0$. To guarantee that all required information is available for the computation of the internal arrival, Step 3 at node $i$ must be executed only after completing Step 2 for all relevant nodes. A straightforward approach to this would be to first complete Step 2 for every node and then move on to Step 3 across all nodes at each time step. More effectively, we can implement the backward Euler approximation through vectorization, which will be elaborated in Section~\ref{sec:simu_vec}.

\begin{figure}[htb]
\begin{center}
\includegraphics[scale=0.3]{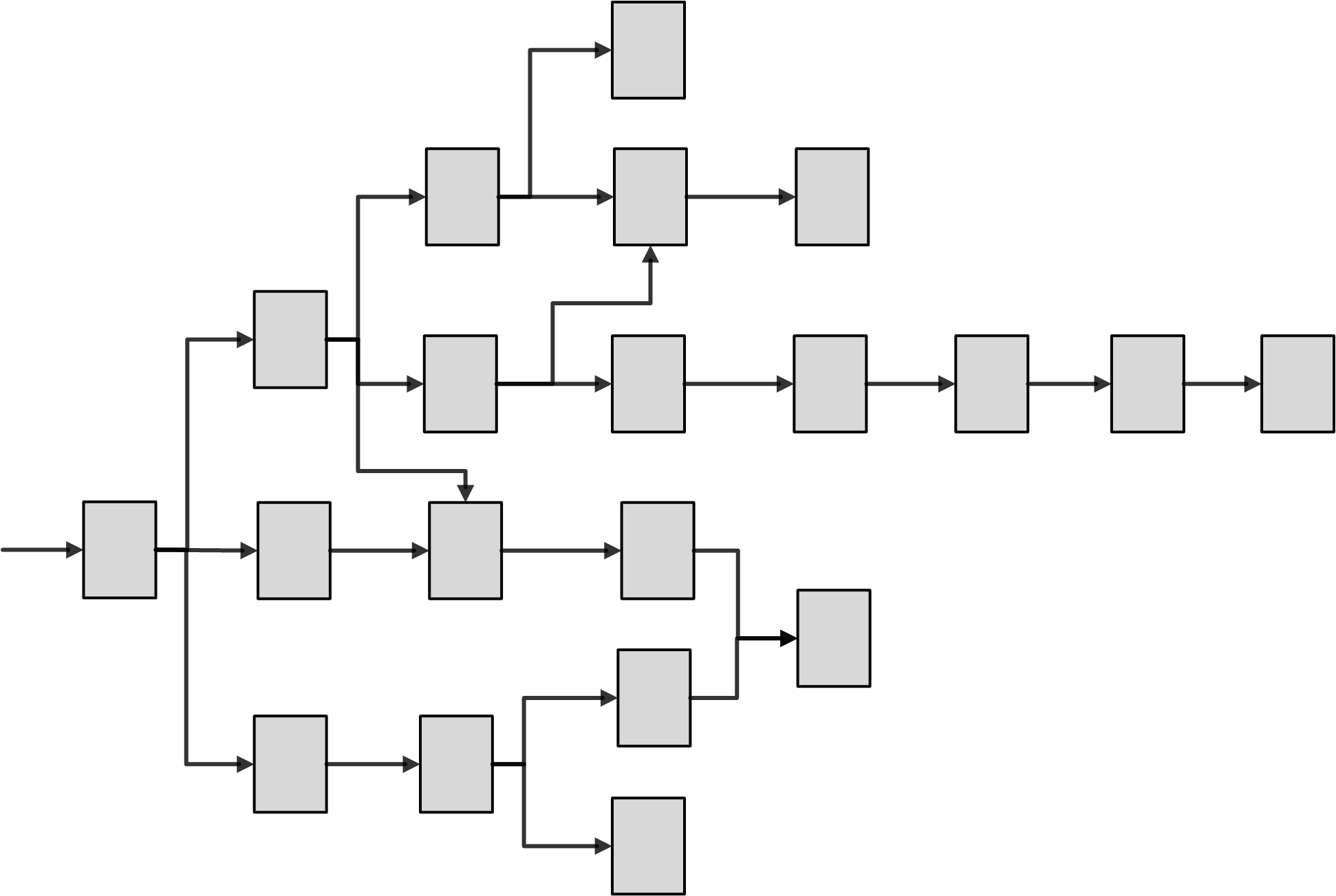}
\caption{An Example of Multi-Layer Feedforward Queueing Network.} \label{fig:multi-layer-net}
\end{center}
\end{figure}

Similarly, we can develop the forward approximation for simulating queueing networks by assuming all the external and internal arrival events occur at the beginning of each time interval. Then the sequence of tasks at
the $i$th node during the $\tau$th time step include:
\begin{enumerate}
\item generate the number of external arrivals  $A^{ex}_{\tau,i}\sim \mathrm{Poisson}\left(\int_{(\tau-1)h}^{\tau h}\lambda_i(u)du\right)$, and compute the number of internal arrivals $A_{\tau,i}^{in,f} = \sum_{k=1}^{n} {{R^f_{\tau,ki}}}$;
\item generate the number of departures $D^f_{\tau,i}\sim\mathrm{GenerateDepartures}(N^f_{\tau-1,i}+A^{ex}_{\tau,i}+A_{\tau,i}^{in,f},m_{\tau,i},\mu_i,h)$;
\item generate $(R^f_{\tau,k1}, \cdots, R^f_{\tau,kn},R^f_{\tau,k\partial},)| D^f_{\tau,k} \sim \mathrm{Multinomial}(D^f_{\tau,k}, (p_{k1},\cdots, p_{kn},p_{k\partial}))$;
\item update the state $N^f_{\tau,i}= N^f_{\tau-1,i}-D^f_{\tau,i}+A^{ex}_{\tau,i}+A^{in,f}_{\tau,i}$.
\end{enumerate}
Notice that implementing a forward scheme poses more challenges compared to a backward scheme. In forward scheme, the computation of internal arrivals at node $i$ in Step 1 is contingent upon the routing outcome from Step 3 at nodes indexed by $k$ for which $p_{ki}>0$. As a result, Step 1 at node $i$ can only proceed after Step 3 has been completed for all nodes that could potentially route customers to node $i$. Additionally, there is a necessity to carry out Steps 1 through 3 in a sequential manner at each individual node. This required sequence of operations introduces significant complexity, particularly when the queueing network contains feedback loops. In the case of a feedforward queueing network, however, it is possible to adopt a forward approximation on a node-by-node basis by aligning with the customer flow direction within the network.  Furthermore, if the queueing network also exhibit multi-layer structure (see Figure~\ref{fig:multi-layer-net} for an example of multi-layer feedforward queueing network), then we can enhance the efficiency of the forward approximation by employing vectorization techniques (see Section~\ref{sec:simu_vec}).

\subsection{Backward and Forward Approximation as the Stochastic Bounds}
Intuitively, the backward scheme assumes the customers are delayed in arriving at the system, and hence they are less likely to finish their service and leave by the end of the time interval. Therefore, the scheme tends to overestimate the true value of system state. Conversely, since the forward scheme assumes the customers arrives earlier, they would have a greater chance of finishing service within the time interval. Consequently, the forward scheme serves as the lower bound for the true value of system state. In Section~\ref{sec:error}, we will present a rigorous proof that the backward and forward Euler approximations provide stochastic upper and lower bounds, respectively, for the exact distribution of the system state in a queueing network. 

For now, we offer a simple illustrative example to clarify the connection between these two Euler schemes and the actual system dynamics. Consider a 2-node tandem queue that is idle at time 0. A customer arrives at the system at time $t_0 \in (0, h)$ and takes service at node 1 for a duration of $s_1$ and at node 2 for $s_2$, with both service times being assumed to be less than $h$. According to the actual system dynamics, this customer would leave the system at time $t_0 + s_1 + s_2\in (0,3h)$. 

\begin{figure}[htb]
{
\centering
\includegraphics[scale=0.41]{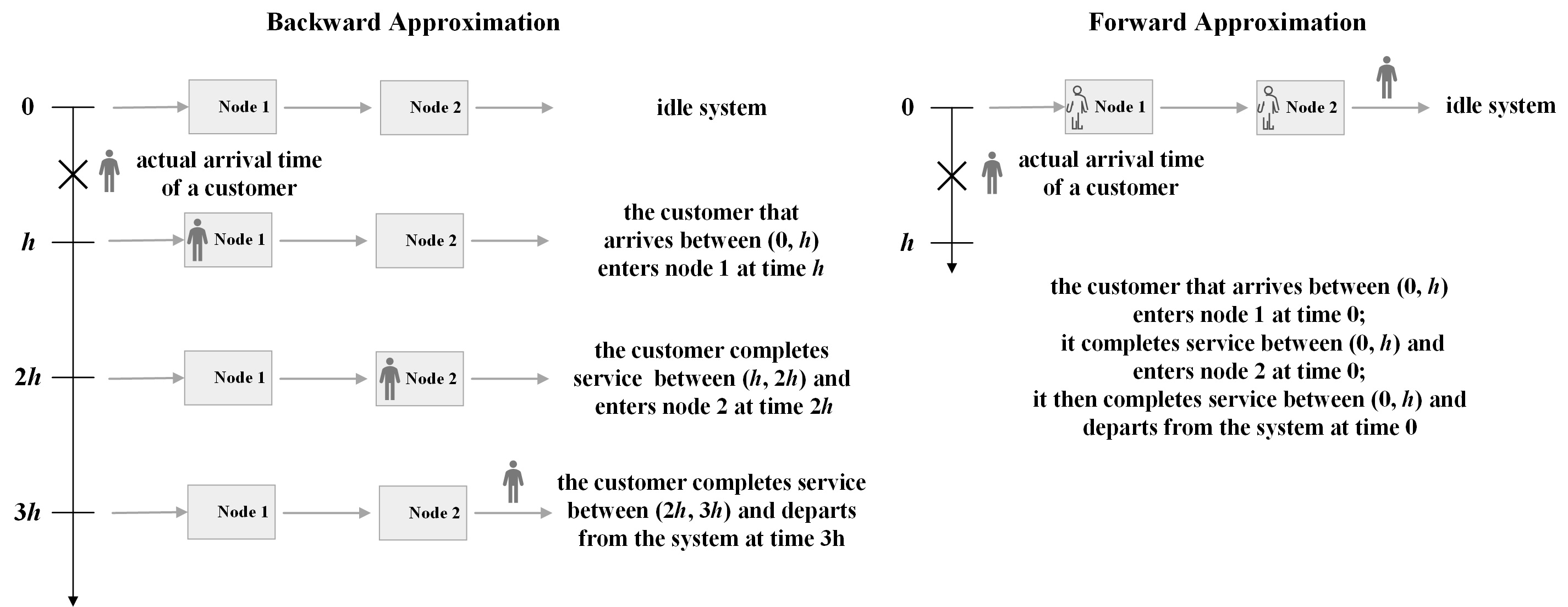}
\caption{Backward Approximation vs. Forward Approximation.\label{fig:backward-forwars}}
}
\end{figure}

Figure \ref{fig:backward-forwars} depicts the trajectory of this customer estimated by the backward approximation scheme (in the left panel) and the forward approximation scheme (in the right panel). With the backward approximation, the customer's entry into nodes 1 and 2 is postponed to times $h$ and $2h$, respectively. Consequently, the customer only finishes service and exits the system just before time $3h$. Counting the number of customers in the system at different time points yields $\left(N_{0}^b, N_{h}^b, N_{2h}^b, N_{3h}^b\right) = (0,1,1,0)$. In contrast, the forward approximation advances the customer's entry into both nodes to immediately after time 0. This means the customer finishes service and leaves the system right instantaneously at time 0. Therefore, we observe $\left(N_{0}^f, N_{h}^f, N_{2h}^f, N_{3h}^f\right) = (0,0,0,0)$. It is evident that the counts $N^b$ and $N^f$ respectively provide upper and lower bounds on the actual state of the system. This example also highlights a fundamental distinction between the two schemes: the backward approximation follows with the natural progression of time, whereas the forward approximation may require backtracking up to one time step to retrieve necessary information.

\subsection{Comparison of Computational Complexity between Euler Approximation and DES}\label{subsec:complexity}

In both backward and forward Euler approximation schemes, we sample the number of arrivals and departures which are aggregated at the interval level, and update the system state based on these aggregated information. Therefore, the simulation cost remains inexpensive even in scenarios with a large number of events in each time interval, which is beneficial especially when dealing with large systems. To demonstrate this point, we can make a comparison on the computational complexity between our Euler approximation approaches and the traditional DES for simulating a large queueing network. For the sake of this comparison, assume the number of servers in all nodes are identical, denoted as $m$, and the service rate of individual server and the utilization of the network are constant as the network size increases.

The computational complexity of DES running over a simulation time horizon of $t$ is known to be ${\rm O}\left( {tmn \log \left( {mn} \right)} \right)$. This stems from the fact that DES processes events sequentially, often utilizing a heap-based data structure to manage these events, which introduces a $\log(mn)$ complexity for event insertion and removal \citep{wang2023fast}. On the other hand, Euler approximation schemes simplify the simulation by dividing time into $t/h$ intervals of fixed length $h$, with the computational effort for each interval being linearly dependent on the number of nodes. Thus, the complexity for Euler approximation over the time horizon $t$ is ${\rm O}\left( tn/h \right)$. Thus, the Euler schemes become more computationally efficient than DES as long as the time step $h$ decreases at a rate slower than ${\rm O}(1/m)$.

The selection of an appropriate time step $h$ is discussed in Section~\ref{sec:selection_rule}. It is demonstrated that our Euler schemes can maintain high accuracy even with relatively large $h$. For example, choosing $h$ on the order of ${\rm O}(1/\sqrt{m})$ allows the relative error in estimating the long-run average number of customers in the system to be asymptotically negligible, while reducing the computational complexity to ${\rm O}\left( tn\sqrt{m} \right)$. This is significantly lower than that of DES essentially when $m$ is large, making the Euler approximation a more scalable approach for simulating large queueing networks.

\section{Vectorized Euler Approximation for Simulating Queueing Networks}\label{sec:simu_vec}

Our approximation methods can achieve even greater efficiency through the use of vectorization techniques, an approach that has been applied successfully to the simulation of large-scale production-inventory systems \citep{wang2023large}. By designing our algorithms in terms of vector and matrix operations, we enable the utilization of parallel computing capabilities inherent in multi-core CPU and many-core GPU architectures, as well as efficient computational tools for vector operations. This approach allows for simultaneous processing of multiple node tasks, significantly accelerating computation and making our methods well-suited for simulating large-scale queueing networks.

In the context of the backward Euler approximation, it is straightforward to employ vectorization to accelerate the computational process. We use boldface notation to denote vectors or matrices. Specifically, the following are definitions of the vectorized quantities:
\begin{equation*}
\setlength{\abovedisplayskip}{10pt}
\setlength{\belowdisplayskip}{10pt}
\begin{aligned}
\boldsymbol{N}^b_{\tau} &= (N^b_{\tau,1}, \cdots, N^b_{\tau,n}),\quad
\boldsymbol{D}^b_{\tau} = (D^b_{\tau,1},\cdots, D^b_{\tau,n}),\\
\boldsymbol{A}^{ex}_{\tau} &= (A^{ex}_{\tau,1},\cdots, A^{ex}_{\tau,n}),\quad
\boldsymbol{A}_{\tau}^{in,b} = (A^{in,b}_{\tau,1},\cdots,A^{in,b}_{\tau,n}),\\
\boldsymbol{m}_{\tau} &= (m_{\tau,1},\cdots,m_{\tau,n}),\quad
\boldsymbol{\mu} = (\mu_1,\cdots, \mu_n),\quad
\mathds{1} = (1,\cdots, 1).
\end{aligned}
\end{equation*}
These quantities are $n$-dimensional row vectors, where $n$ is the number of nodes in the network. For the matrices:
\begin{equation*}
\setlength{\abovedisplayskip}{7pt}
\setlength{\belowdisplayskip}{7pt}
\begin{aligned}
\boldsymbol{P} = (p_{ij})_{n\times n},\quad
\boldsymbol{R}_\tau^b = (R_{\tau, ij}^b)_{n\times n}.
\end{aligned}
\end{equation*}
Additionally, $\boldsymbol{P}_i$ and $\boldsymbol{R}^b_{\tau,i}$ denote the $i$th row of the respective matrices.
In real-world applications, the routing matrix $\boldsymbol{P}$ (and hence $\boldsymbol{R}^b_\tau$) often has many zero entries. Under such circumstances, we can use sparse matrix techniques to efficiently store and handle these matrices, which can significantly reduce memory requirements and computational time.

Algorithm~\ref{algo:backward-vectorized} presents the vectorized version of backward approximation scheme for simulating queueing networks. It involves a vectorized version of the algorithm $\mathrm{GenerateDepartures}$, which is provided in Appendix~\ref{sec:pure_departure_vec}.
These algorithms leverage the ability to generate multivariate random variables through vectorized operations, a feature supported by many computational software packages, including MATLAB and Python library NumPy. These software environments are highly optimized for matrix and vector operations, which can lead to substantial performance gains.

\SetAlgoSkip{medskip}
\begin{algorithm}[H]
  \caption{Vectorized Backward Euler Approximation for Queueing Networks}\label{algo:backward-vectorized}
  
  initialization of $\boldsymbol{N}^b_{0}$, $\boldsymbol{m}_{\tau}$, $\boldsymbol{\mu}$, $h$\;\vspace{5pt}
  
  \For{$\tau = 1$ \KwTo $t/h$}{
  \vspace{5pt}
   generate $\boldsymbol{D}^b_{\tau}\sim\mathrm{GenerateDepartures}(\boldsymbol{N}^b_{\tau-1},\boldsymbol{m}_{\tau},\boldsymbol{\mu},h)$\;\vspace{5pt}
   generate $\boldsymbol{R}^b_\tau$, where the $i$th row $ \boldsymbol{R}^b_{\tau,i} \sim \mathrm{Multinomial}(\boldsymbol{D}^b_{\tau,i}, {\boldsymbol{P}_i})$\;\vspace{5pt}
   generate the number of external arrivals  $\boldsymbol{A}^{ex}_{\tau}$\;\vspace{5pt}
   compute the number of internal arrivals $\boldsymbol{A}_{\tau}^{in,b} = \mathds{1} \boldsymbol{R}^b_\tau $\;\vspace{5pt}
   update the state $\boldsymbol{N}^b_{\tau}= \boldsymbol{N}^b_{\tau-1}-\boldsymbol{D}^b_{\tau}+\boldsymbol{A}^{ex}_{\tau}+\boldsymbol{A}^{in,b}_{\tau}$\;
 }
\end{algorithm}

In contrast to the backward approximation which allows for comprehensive vectorization, the forward approximation's potential for vectorization is limited due to the inter-dependencies of node states within a given time interval: the number of arrivals at the $i$th node depends on the departures from preceding nodes, which in turn are influenced by their respective arrivals. However, it is possible to achieve partial vectorization with the forward approximation in the context of a multi-layer feedforward queueing network. In such a network, nodes are organized into several sequential layers, and customers progress strictly from a node in one layer to nodes in the next. This structure permits the concurrent processing of nodes within the same layer, thereby improving computational efficiency. 

Algorithm~\ref{algo:forward-vectorized} presents the vectorized implementation of the forward Euler approximation scheme for a multi-layer queueing network with $l$ layers. Here $\mathcal L_j$ denotes the index set of nodes in the $j$th layer for $j=1,\cdots, l$.  Consistent with the notations for the backward scheme, we use boldface notation to denote vectors or matrices. When a vector is subscripted with $\mathcal L_j$, this indicates that the vector's components are comprised of values associated with nodes whose indices are in $\mathcal L_j$. For example, $\boldsymbol{N}^f_{\tau,\mathcal L_j}$ represents a row vector that contains the elements ${N^f_{\tau,i}, i \in \mathcal L_j}$. $\boldsymbol{R}^f_{\tau,{\mathcal L_j}}$ represents a matrix with entries $R^f_{\tau,ik}$ with $i\in\mathcal L_j$ and $k\in\mathcal L_{j+1}$.

\SetAlgoSkip{medskip}
\begin{algorithm}[H]
  \caption{Vectorized Forward Euler Approximation for Multi-layer Queueing Networks}\label{algo:forward-vectorized}
  initialization of $\boldsymbol{N}^f_{0}$, $\boldsymbol{m}_{\tau}$, $\boldsymbol{\mu}$, $h$ and $\boldsymbol{R}^f_{\tau,{\mathcal L_0}}=\boldsymbol{0}$\;\vspace{5pt}
  
  \For{$\tau = 1$ \KwTo $t/h$}{
  \vspace{5pt}
   generate the number of external arrivals  $\boldsymbol{A}^{ex}_{\tau}$\;\vspace{5pt}
  \For{$j=1$ \KwTo $l$}{
  \vspace{5pt}
   compute the number of internal arrivals $\boldsymbol{A}_{\tau,{\mathcal L_j}}^{in,f} = \mathds{1} \boldsymbol{R}^f_{\tau,{\mathcal L_{j-1}}}$\;\vspace{5pt}
   generate $\boldsymbol{D}^f_{\tau,{\mathcal L_j}}\sim\mathrm{GenerateDepartures}(\boldsymbol{N}^f_{\tau-1,{\mathcal L_j}}+ \boldsymbol{A}_{\tau ,{\mathcal L_j}}^{ex}+ \boldsymbol{A}_{\tau ,{\mathcal L_j}}^{in,f},\boldsymbol{m}_{\tau,{\mathcal L_j}},\boldsymbol{\mu}_{{\mathcal L_j}},h)$\;\vspace{5pt}
   update the state $\boldsymbol{N}^f_{\tau,{\mathcal L_j}}= \boldsymbol{N}^f_{\tau-1,{\mathcal L_j}}-\boldsymbol{D}^f_{\tau,{\mathcal L_j}}+\boldsymbol{A}^{ex}_{\tau,{\mathcal L_j}}+\boldsymbol{A}^{in,f}_{\tau,{\mathcal L_j}}$\;\vspace{5pt}
   generate $\boldsymbol{R}^f_{\tau,{\mathcal L_j}}$, where the $i$th row ($i\in\mathcal L_j$) $\boldsymbol{R}^f_{\tau,i} \sim \mathrm{Multinomial}(\boldsymbol{D}^f_{\tau,i}, {\boldsymbol{P}_i})$\;
 }
 }
\end{algorithm}

Compared with Algorithm \ref{algo:backward-vectorized}, Algorithm \ref{algo:forward-vectorized} incorporates an additional loop to ensure that the simulation progresses according to the sequential order of the layers. Due to this layer-by-layer progression, the forward approximation may not fully exploit the potential acceleration
that vectorization offers compared to the backward approximation. Therefore, although both the node-by-node versions of the backward and forward approximations have the same computational complexity as discussed in Section~\ref{subsec:complexity}, the vectorized version of the backward approximation can be more efficient in practice.

\section{Pure Departure Process}\label{sec:pure_departure}

A pure departure process refers to the queueing system dynamics where current customers depart after receiving service without any new customers arriving. It plays a pivotal role in our proposed simulation approaches, both in terms of the algorithms and the theoretical results related to the algorithms. As detailed in Section~\ref{sec:sim_state}, to simulate single-station queues and queueing networks with the Euler approximation, it requires the simulation of the number of departures for a pure departure process. Section~\ref{sec:pure_departure_simu} examines various simulation strategies for the pure departure process and introduces a scheme that can simulate it exactly. Additionally, Section~\ref{sec:error_pre} elaborates on the properties of the pure departure process, laying the groundwork for subsequent analyses in Sections~\ref{sec:error} and \ref{sec:sojourn}.

\subsection{Simulation of Pure Departure Process}\label{sec:pure_departure_simu}

In this subsection, we examine the simulation of a pure departure process from a queueing system equipped with $m$ identical servers, each operating at an exponential service rate of $\mu$. We assume the system starts with $x$ customers at time 0, and we focus on simulating $D$, which represents the stochastic count of customers who complete service within the time interval $h$. The value of $D$ depends on the overall service capacity, which is determined by the number of busy servers.
Let $B_t$ denote the number of busy servers at time $t$.  It is
critical to note that $\{B_t,0\leq t\leq h\}$ is a stochastic process taking values in $\{0,1,\cdots,m\}$ and it is non-increasing for the pure departure process.

\subsubsection{Na\"ive Euler Approximation} By the nature of Euler approximation, we may approximate $B_t$ by $b=B_0 = \min\{x, m\}$ for any $t\in[0, h)$. As a result, we may set
\begin{equation}\label{eq:pure_simu_1}
\setlength{\abovedisplayskip}{3pt}
\setlength{\belowdisplayskip}{3pt}
D =\min\{x,S\}, \quad \text{where } S\sim \mathrm{Poisson}(b\mu h).
\end{equation}
Here $S$ denotes the overall service capacity within the time step, and the minimum operator ensures that the number of customers in the system does not fall below zero. Notice that this scheme may over-estimate the number of busy servers, as $B_t$ is non-increasing in $(0, h)$ thus may drop below $b$. As a consequence, there is a tendency for this approximation to overestimate $D$.

\subsubsection{A Simple Refinement} It is clear to observe that when $x\leq m$, each sever can serve at most one customer. This means the completion of service by a server will reduce $B_t$ by one. Therefore, in this case, $D$ follows a binomial distribution with parameters $x$ and $p=1-e^{-\mu h}$, where $p$ is the probability that a busy server completes service within $(0, h)$. Thus, we may set
\begin{equation}\label{eq:pure_simu_2}
\setlength{\abovedisplayskip}{10pt}
\setlength{\belowdisplayskip}{10pt}
\begin{aligned}
D &=\min\{x,S\}\cdot 1_{\{x\geq m+1\}} + \tilde S\cdot 1_{\{x\leq m\}}, \\
&\text{where } S\sim \mathrm{Poisson}(b\mu h),\quad \tilde S\sim \mathrm{Binomial}\left(x,p\right).
\end{aligned}
\end{equation}
Notice that Equation~\eqref{eq:pure_simu_2} refines Equation~\eqref{eq:pure_simu_1} by providing an exact simulation for the case where $x\leq m$. Nevertheless, this refined approach still overestimates $D$ when $x \geq m + 1$.

\subsubsection{The Exact Simulation} Let $T$ be the time at which the $(x - m)$th customer completes
service when $x \geq m + 1$. It is critical to recognize that in this case, $B_t = m$ for all $t \in [0, T)$. Therefore, $T$ follows $\mathrm{Erlang}(x - m, m\mu)$. Conditioning on the value of $T$, the service completion times for the $x - m$ customers are uniformly distributed over $[0, T]$, and hence each customer leaves the system during $[0,h)$ independently with probability $h/T$ if $T>h$.
Furthermore, if $T < h$, then at $t = T$, we have exactly $m$ customers in the system, reducing the situation to the previous case where $x\leq m$ considered in the simple refinement, but now with the remaining time $h - T$.  This allows for the exact characterization of the distribution of $D$, which is summarized by the following lemma (the proof can be found in Appendix~\ref{proof:pure-departure-1}):

\begin{lemma}\label{lem:pure-departure-1}
Consider the pure departure process with $m$ servers, each having a service rate of $\mu$. Assume initially there are $x$ customers in the system. Let $D$ be the random number of customers completing service by time $h$. 
\begin{enumerate}
\item[(1)] If $x\leq m$, then $D$ follows the binomial distribution $\mathrm{Binomial}(x, 1-e^{-\mu h})$.
\item[(2)] If $x\geq m+1$, suppose the $(x-m)$th departing customer leaves the system after a time period of $T$, then $T$ follows  $\mathrm{Erlang}(x-m,m\mu)$. Conditioning on $T$,
\begin{enumerate}
\item[(2.1)] when $T>h$, then the conditional distribution of $D$ follows $\mathrm{Binomial}(x-m-1,h/T)$;
\item[(2.2)] when $T\leq h$, then the conditional distribution of $D-x+m$ follows $\mathrm{Binomial}(m, 1-e^{-\mu \left(h-T\right)})$.
\end{enumerate}
\end{enumerate}
\end{lemma}

Algorithm~\ref{alg:pure-departure} is designed based on the above lemma, and thus serves as an exact simulation scheme for $D$.  The merits of this exact simulation scheme are twofold. On the one hand, as the simulation of the pure departure process is exact, the error of our Euler approximation schemes comes solely from the aggregation of arrivals. It enables us to conduct an in-depth error analysis and derive a theoretical error bound (see Section~\ref{sec:error} for details). On the other hand, this sampling scheme is efficient. In contrast to the DES method, it doesn't rely on the event sequence and doesn't involve any iterations. As a result, this scheme maintains a constant-order computational complexity. This property is particularly beneficial when applying our approach to the simulation of large-scale queueing networks.

\SetAlgoSkip{medskip}
\begin{algorithm}[H]
\caption{Generate Departure for a Pure Departure Process}\label{alg:pure-departure}
\KwIn{initial state $x$, number of servers $m$, service rate $\mu$, interval length $h$}\vspace{5pt}
\KwOut{a random sample for the number of departures $D$}\vspace{5pt}

\uIf{$x\leq m$}{sample $D$ from $\mathrm{Binomial}(x, 1-e^{-\mu h})$\;}
\uElse{
sample $T$ from $\mathrm{Erlang}(x-m,m\mu)$\;\vspace{5pt}
\uIf{$T>h$}{sample $D$ from $\mathrm{Binomial}(x-m-1,h/T)$\;}
\uElse{sample $D'$ from $\mathrm{Binomial}(m, 1-e^{-\mu \left(h-T\right)})$\;\vspace{5pt}
set $D=D'+x-m$.}
}
\end{algorithm}

\subsection{Properties of the Pure Departure Process}\label{sec:error_pre}

This subsection outlines some properties of the pure departure process that will be used in subsequent sections. In particular, Lemma~\ref{lem:pure-departure-2} establishes the stochastic dominance properties of the pure departure process, which form the basis for the error analysis discussed in Section~\ref{sec:error}. We use the notation $X\preceq Y$ to denote that the random variable $Y$ has \emph{the first order stochastic dominance} over random variable $X$. This means that for all values $x$, the probability that $X$ is greater than or equal to 
$x$ is less than or equal to the probability that $Y$ is greater than or equal to $x$, i.e., $P(X\geq x)\leq P(Y\geq x)$. When $X\preceq Y$, it is said that $X$ is a stochastic lower bound for $Y$, and $Y$ is a stochastic upper bound for $X$.

\begin{lemma}\label{lem:pure-departure-2}
For given $t\geq0$ and non-negative integer $x\geq0$, define $\mathcal G_t(x)$ to be a non-negative integer-valued random variable, of which the distribution is specified by
\[
\setlength{\abovedisplayskip}{3pt}
\setlength{\belowdisplayskip}{3pt}
\mathcal G_t(x)\eqd x - D, \quad \text{with } D\sim \mathrm{GenerateDeparture}(x, m, \mu, t),
\]
where ``$\eqd$" means identical in distribution. Suppose $X$ and $Y$ are non-negative integer-valued random variables, then
\begin{enumerate}
\item[(i)] if $X\preceq Y$, then $\mathcal G_t(X)\preceq \mathcal G_t(Y)$;
\item[(ii)] for any $t,s\geq 0$, $\mathcal G_{t+s}(X+1)\preceq \mathcal G_t(\mathcal G_s(X)+1)\preceq \mathcal G_{t+s}(X)+1$.
\end{enumerate}
\end{lemma}

Intuitively, $\mathcal G_t(x)$ represents the distribution of the number of customers in a pure departure system at time $t$, given that there are initially $x$ customers at time 0. Property (i) states that an increase in the stochastic order of the initial customer count in a pure departure process will result in a corresponding increase in the stochastic order of the customer count at the end of the process. In property (ii), $\mathcal G_t(\mathcal G_s(X)+1)$ represents the distribution of the number of customers at time $t+s$, assuming we start with $X$ customers at time 0 and exactly one customer arrives at time $s$. This property suggests that moving the arrival time of this customer to the beginning of the period (time 0) decreases the stochastic order of the terminal customer count, whereas delaying this arrival to the end of the period (time $t+s$) increases the stochastic order.  It carries the essential idea of our forward/backward Euler approximation scheme, and it is crucial for the proof of Theorem~\ref{thm:error}.
The proof of the lemma can be found in Appendix~\ref{proof:pure-departure-2}.

Lemma~\ref{lem:pure-departure-3} elucidates a critical aspect of pure departure process that is instrumental in developing our simulation algorithm for sojourn times, as detailed in Section~\ref{sec:sojourn}. This lemma quantifies the probability that a chosen server, initially serving a customer, continues to serve that same customer after a predetermined time period, given the total number of service completions during that period. The insights derived from this lemma allow for a conditional simulation approach to sample the time a customer spends in the system. The proof of the lemma can be found in Appendix~\ref{proof:pure-departure-3}.
\begin{lemma}\label{lem:pure-departure-3}
Consider a pure departure process with $m$ servers, each having a service rate of $\mu$. Assume initially there are $x\geq 1$ customers in the system.  Select a server that is actively servicing a customer at the start of the observation, and denote by $\mathcal E$ the event that this server continues to serve the same customer after a specified time interval. Let $D$ be the total number of customers who complete service during this time interval. 
\begin{enumerate}
\item[(1)] If $x\leq m$, then
\[
P\left(\mathcal E \Big| D = d\right) = \frac{x-d}{x}, \text{ for } d = 0, 1, \cdots, x.
\]
\item[(2)] If $x\geq m+1$, then
\[
P\left(\mathcal E \Big| D = d\right) = 
 \begin{cases}
      \left(\dfrac{m-1}{m}\right)^d, & \text{if } d = 0, 1, \cdots, x-m;\\
      \left(\dfrac{m - 1}{m}\right)^{x-m}\dfrac{x - d}{m}, & \text{if } d = x-m, \cdots, x.
 \end{cases}  
\]
\end{enumerate}
\end{lemma} 

\section{Analysis of the Approximation Errors}\label{sec:error}
This section focuses on analyzing the performance of the backward and forward approximations developed in Section~\ref{sec:sim_state}. Our primary objective is to establish the relationship between the choice of the step size of time discretization and the resulting approximation errors. We first consider the case for a single-station queue and conduct the transient analysis in Subsection~\ref{sec:error_single}, and conduct the steady-state analysis for queueing networks in Subsection~\ref{sec:error_network}. Then in Subsection~\ref{subsec:asymptotic_error} we consider the asymptotic error analysis for simulating large systems, and discuss the selection rule for step size $h$.

\subsection{Transient Analysis for the Single-Station Queue}\label{sec:error_single}
In this subsection, we focus on the error analysis of our approach for simulating the single-station queue. The configuration of the single-station queue is described by the following assumption.
\begin{assumption}\label{asm:single}
The single-station queue is modeled by a multi-server Markovian queue with $m$ servers and no limit of buffer. The arrival rate $\lambda$ and the service rate $\mu$ are both positive constants. The system is initially idle.
\end{assumption}

We aim to demonstrate that the two Euler approximates, $N_\tau^b$ and $N_\tau^f$, generated respectively by the backward and forward schemes, are close in distribution to the true system state $N_\tau$ at time $t=\tau h$ for any $\tau\geq 1$. Our error analysis progresses through three fundamental arguments, forming the basis for establishing the main results outlined in Theorem~\ref{thm:error}. The comprehensive and rigorous analysis based on these arguments is presented in the proof of the theorem.

\begin{argument} $N_\tau^f\preceq  N_\tau\preceq  N_\tau^b$ for any $\tau\geq0$.
\end{argument}

This argument says that the forward and backward schemes generate stochastic lower bound and upper bound for the number of customers in the single-station queue, respectively. It aligns with intuition. Since the forward scheme advances the customer arrival times to the beginning of each time interval, it can make better use of the service capacities and reduce the number of customers in the station. Conversely, the backward scheme delays the entry of the arriving customers to the end of each time interval, depriving them of service completion within the time interval, and hence all the new arrivals will be included in the system state counts.

We claim that the upper bound and lower bound hold rigorously in the sense of first order stochastic dominance. The key observation here is that the distribution of $N_\tau$, $N_\tau^b$ and $N_\tau^f$ can be represented in terms of $\mathcal G_h$ through conditional probabilities. In particular, according to the sampling rule of the approximation schemes detailed in Section~\ref{subsubsec:sim_N}, we have
\begin{equation*}
\setlength{\abovedisplayskip}{10pt}
\setlength{\belowdisplayskip}{10pt}
\begin{aligned}
  N_\tau^b | \{N_{\tau-1}^b;A_\tau\} &\eqd \mathcal G_{h}\left(N_{\tau-1}^b\right) + A_\tau,\\
  N_\tau^f | \{N_{\tau-1}^f;A_\tau\} &\eqd \mathcal G_{h}\left(N_{\tau-1}^f+ A_\tau\right),
\end{aligned}
\end{equation*}
where $A_\tau$ is the aggregated number of arrivals within the $\tau$th time interval. 
Moreover, conditional on $A_\tau = k$, and that the $k$ arrival times are $T_1, T_2, \cdots, T_{k}$ ($(\tau-1)h<T_1<\cdots<T_{k}<\tau h$), the evolution of the system state between any two consecutive arrival times follows the pure departure process. This allows us to express
\begin{equation*}
\setlength{\abovedisplayskip}{10pt}
\setlength{\belowdisplayskip}{10pt}
\begin{aligned}
~&N_\tau | \{N_{\tau-1};A_\tau=k; T_1, T_2, \cdots, T_{k}\}\\
~&\eqd  \mathcal G_{\tau h-T_{k}}\left(1+\mathcal G_{T_{k}-T_{k-1}}\left(\cdots1+\mathcal G_{T_2-T_1}\left(1+ \mathcal G_{T_1-(\tau-1) h} (N_\tau)\right)\cdots\right)\right).
\end{aligned}
\end{equation*}
By induction on $\tau$ and employing Lemma~\ref{lem:pure-departure-2} (i) and (ii) repeatedly,
we can prove the stochastic orders of the three random variables based on the above representations.

\begin{argument} $(N_{\tau}^b,D_{\tau+1}^b)$ and $(N_{\tau-1}^f +A_{\tau}, D_{\tau}^f)$ are identical in distribution for any $\tau\geq 1$.
\end{argument}

Again, we establish the argument by induction on $\tau$. It is straightforward to verify the argument for $\tau=1$.
Assume we have established that $(N_{\tau-1}^b, D_\tau^b)$  and $ (N_{\tau-2}^f +A_{\tau-1}, D_{\tau-1}^f)$ are identical in distribution for a given $\tau$, then 
$$
\setlength{\abovedisplayskip}{3pt}
\setlength{\belowdisplayskip}{3pt}
N_{\tau-1}^b-D_\tau^b \eqd N_{\tau-2}^f +A_{\tau-1}- D_{\tau-1}^f.$$ 
Furthermore, notice that all random variables in this identity only depend on the arrivals up to time $(\tau-1)h$ and hence are independent with $A_\tau$, therefore if we add $A_\tau$ on both sides, the identity still holds. Combining with the updating equations for backward and forward schemes, we have
\begin{equation*}
\setlength{\abovedisplayskip}{6pt}
\setlength{\belowdisplayskip}{6pt}
\begin{aligned}
    N_\tau^b &= N_{\tau-1}^b-D_\tau^b+A_\tau
    \eqd N_{\tau-2}^f +A_{\tau-1} - D_{\tau-1}^f+A_\tau
    =N_{\tau-1}^f+A_{\tau}.
\end{aligned}
\end{equation*}
Then according to the sampling rule in Section~\ref{subsubsec:sim_N}, we have 
\begin{equation*}
\setlength{\abovedisplayskip}{9pt}
\setlength{\belowdisplayskip}{9pt}
\begin{aligned}
 D_{\tau+1}^b&\sim \mathrm{GenerateDeparture}(N_{\tau}^b, m, \mu, h),\\
 D_\tau^f&\sim \mathrm{GenerateDeparture}(N_{\tau-1}^f+A_\tau, m, \mu, h).
\end{aligned}
\end{equation*}

It suggests that the distribution of $D_{\tau+1}^b$ conditional on $N_\tau^b=x$ is identical to the distribution of $D_\tau^f$ conditional on $N_{\tau-1}^f+A_\tau=x$, for any non-negative integer $x$. Therefore, we have $(N_{\tau}^b,D_{\tau+1}^b) \eqd (N_{\tau-1}^f +A_{\tau}, D_{\tau}^f)$.

\begin{argument} $E[N_\tau^b]-E[N_\tau^f] =E[D_\tau^f]$, for any $\tau\geq0$.
\end{argument}

Since $N_\tau^b$ and $N_{\tau-1}^f + A_\tau$ are identical in distribution, we have
$$
\setlength{\abovedisplayskip}{3pt}
\setlength{\belowdisplayskip}{3pt}
E[N_\tau^b]-E[N_\tau^f]=E[N_{\tau-1}^f + A_\tau]-E[N_\tau^f]=E[N_{\tau-1}^f + A_\tau-N_\tau^f]=E[D_\tau^f].$$
Because the service rate of the station is at most $m\mu$, we can further conclude that 
$E[N_\tau^b]-E[N_\tau^f]\leq m\mu h$. Moreover, in steady state, the average number of departures from the station equals the average number of arrivals to the station, so we have $E[N^b]-E[N^f] = \lambda h$.

Based on the above arguments, we can prove the following theorem rigorously. The detailed proof is deferred to Appendix~\ref{proof:error}.

\begin{theorem}\label{thm:error}
Consider the simulation of a single-station queue satisfying Assumption~\ref{asm:single}. For fixed $h>0$, let $N_\tau$ be the number of customers in the system at time $t=\tau h$, and $N_\tau^b$ and $N_\tau^f$ be the number of customers generated by the backward and forward approximation schemes. Then we have
\[
\setlength{\abovedisplayskip}{3pt}
\setlength{\belowdisplayskip}{3pt}
N_\tau^f\preceq  N_\tau\preceq  N_\tau^b,\quad E[N_\tau^b]-E[N_\tau^f] \leq m\mu h.
\]
Furthermore, when $\lambda<m\mu$, let $N$, $N^b$ and $N^f$ are the steady-state versions of $N_\tau$, $N_\tau^b$ and $N_\tau^f$, then we have
\[
\setlength{\abovedisplayskip}{3pt}
\setlength{\belowdisplayskip}{3pt}
N^f\preceq  N\preceq  N^b,\quad E[N^b]-E[N^f] = \lambda h.\]
\end{theorem}

Theorem~\ref{thm:error} shows that the errors associated with forward and backward Euler approximation schemes decay linearly with the step size $h$, consistent with the traditional Euler scheme used for simulating diffusion processes. Interestingly, the error bound does not depend on $\tau$, indicating that the magnitude of the approximation error does not propagate and accumulate over the simulation horizon. This characteristic of uniform convergence over time sets it apart from the typical behavior of Euler schemes for diffusion processes, where uniform convergence over time is not guaranteed. The phenomenon of uniform convergence in our context is attributed to the inherent ``self-correction" feature of our approximation schemes for queueing models.  In the backward scheme, where the number of customers is stochastically higher, the system's overall service rate increases correspondingly, serving as a corrective force that realigns the state of the system closer to its true value. Conversely, in the forward scheme, a stochastically lower customer count leads to a reduced service rate, serving as an upward corrective pressure that pushes the system state approximation towards its true value. Consequently, throughout the simulation, regardless of 
$\tau$, the discrepancy between the upper and lower bounds of the expected customer count in the system remains consistently capped at $m\mu h$. When the system reaches equilibrium, this gap converges to $\lambda h$, reinforcing the robustness of the approximation schemes in capturing the system's dynamics over time.

\subsection{Steady-State Analysis for the Feedforward Queueing Network}\label{sec:error_network}
Now we extend our discussion from the single-station queue to the queueing networks.
The results in this subsection are established under the following assumption:

\begin{assumption}\label{asm:network}
The queueing network consists of $n$ interconnected nodes, where the $i$th node is modeled by a single-station queue with $m_i$ servers, each server operating at an exponential service rate of $\mu_i$. The network is feedforward, meaning that once customers complete service at one node and leave, they do not return to the same node. The transition of customers among nodes is dictated by the routing matrix $\mathbf{P}=(p_{ij})_{n\times n}$. Customers from outside of the system arrive at node $i$ at a rate of $\lambda_i$.%
\end{assumption}

Although the queueing network can be viewed as a group of interconnected single-station queues, the application of the Euler approximation schemes to the queueing network introduces unique challenges in error analysis that are not encountered with single-station queues. To illustrate the problem, let's consider a two-station tandem system. In the backward approximation, as we shift all the external arrivals in the first time interval backward to time $h$, the counts of system state at time $h$ for both two stations tend to be stochastically larger because the arrival of external customers is delayed. However, the delay at the first station results in a decrease in the number of customers arriving at the second station during the interval. Indeed, in this particular example, the second station would experience no internal arrivals within the first interval, as the delayed customers from the first station have not yet had the opportunity to transition to the second station. As a result, it tends to reduce the level of system state for the second station. This exemplifies the intricate interplay within a multi-station network and highlights why transient analysis for single-station queues does not directly extend to networked systems.

Alternatively, we argue that the stochastic order between the backward/forward approximations and the true system dynamics remains valid for feedforward queueing networks in steady state. We present the result in the following theorem, and the proof is provided in Appendix~\ref{proof:error_network}.

\begin{theorem}\label{thm:error_network}
Consider the simulation of a queueing network satisfying Assumption~\ref{asm:network}. In addition, we assume the utilization of every node in the network is less than 1. 
\begin{enumerate}
\item[(i)] Let $N(i)$ be the total number of customers in the $i$th node in steady state distribution, and $N^b(i)$ and $N^f(i)$ be the corresponding total number of customers generated by the backward and forward approximation schemes with step size $h>0$. Then
$$
\setlength{\abovedisplayskip}{3pt}
\setlength{\belowdisplayskip}{3pt}
N^f(i)\preceq N(i)\preceq N^b(i), \quad E[N^b(i)]-E[N^f(i)] = \tilde \lambda_i h,$$
where $\tilde \lambda_i$ is the total arrival rate to the $i$th node satisfying
$\tilde \lambda_i=\lambda_i + \sum_{k=1}^n p_{ki}\tilde\lambda_k$, $i=1,\ldots,n.$
\item[(ii)] Let $N$ be the total number of customers in the system in steady state distribution, and $N^b$ and $N^f$ be the corresponding total number of customers generated by the backward and forward approximation schemes with step size $h>0$. Then 
\[
\setlength{\abovedisplayskip}{3pt}
\setlength{\belowdisplayskip}{3pt}
N^f\preceq N\preceq N^b,\quad E[N^b]-E[N^f] = \sum_{i=1}^n \tilde \lambda_i h.\]

Furthermore, if the network is of multi-layer structure and has $l$ layers, then
\begin{equation}
\setlength{\abovedisplayskip}{5pt}
\setlength{\belowdisplayskip}{5pt}
E[N^b]-E[N^f] \leq \sum_{k=1}^l \Lambda_k(l+1-k) h ,\label{eq:gap_network}
\end{equation}
where $\Lambda_k$ is the total external arrival rate to the $k$th layer.
\end{enumerate}
\end{theorem}

As an example, consider the scenario when the external customers enter the system only through the first-layer nodes. Then the right-hand side of \eqref{eq:gap_network} becomes $\Lambda l h$, where $\Lambda$ is the total external arrival rate. This result can also be derived using Little's law. For either backward or forward Euler approximation scheme, whenever a customer enters a new layer, it will incur a discretization error of at most $h$ to the time that this customer spends in the system. Therefore, after the customer goes through all $l$ layers, the total error on the time that this customer spends in the system is at most $lh$. Then applying Little's law, we can conclude that the approximation error on the expected number of customers in the system is at most $\Lambda l h$. In general, the error bound in \eqref{eq:gap_network} is smaller than $\Lambda l h$, as the external arrivals to the $k$th layer only go through $l+1-k$ layers.

\subsection{Asymptotic Analysis on Relative Errors for Simulating Large Systems}\label{subsec:asymptotic_error}

In this section, we focus on the error analysis for Euler approximation methods when applying to the simulation of large-scale queueing networks. We consider the asymptotic regimes where both the number of servers and the system utilization are increasing, while the service rate per individual server and the network's structure remain unchanged. This scenario aligns with common real-world circumstances where, to maintain a balanced system load, the growth in arrival rates is matched by an increase in the number of staffed servers. 
It is important to note that performance measures in a large queueing network are more meaningful when they are scaled to the size of the system. Relative errors, which are expressed as a proportion of the total, automatically scale with the size of the network and provide a more accurate reflection of the system's performance.  Consequently, our investigation will focus on the behavior of relative errors of our approximation methods as the network scales up. This will provide a clearer understanding of the scalability and reliability of simulation outputs, which are critical for effective network management and optimization.

\subsubsection{Single-Station Queue}

Consider a single-station queue satisfying Assumption~\ref{asm:single}, and we aim to evaluate the long-run average number of customers in the system through Euler approximation schemes. Using the same notations in Theorem~\ref{thm:error}, we define the relative errors for backward and forward schemes by
\[
\mathrm{RE}^b(h) = \frac{|E[N^b]-E[N]|}{E[N]}\quad\text{and}\quad \mathrm{RE}^f(h) = \frac{|E[N^f]-E[N]|}{E[N]}.
\]
Furthermore, we introduce the term
\[
\overline{\mathrm{RE}}(h) = \frac{\lambda h}{E[N]}.
\]
Based on the results of Theorem~\ref{thm:error}, we know that
\[
\setlength{\abovedisplayskip}{3pt}
\setlength{\belowdisplayskip}{3pt}
\mathrm{RE}^b(h) \leq \overline{\mathrm{RE}}(h),\quad \mathrm{RE}^f(h) \leq \overline{\mathrm{RE}}(h),
\]
and
\[
\setlength{\abovedisplayskip}{5pt}
\setlength{\belowdisplayskip}{5pt}
\mathrm{RE}^b(h) + \mathrm{RE}^f(h) = \overline{\mathrm{RE}}(h).
\]
Thus, $\overline{\mathrm{RE}}(h)$ serves not just as an upper bound for the relative errors of both Euler schemes but also accurately reflects the asymptotic magnitude of these errors. This allows us to investigate the asymptotic behavior of the relative errors by examining the properties of $\overline{\mathrm{RE}}(h)$.  Our findings are summarized in Theorem~\ref{thm:relative_error}, with the proof being provided in Appendix~\ref{proof:relative_error}.

\begin{theorem}\label{thm:relative_error}
Consider the simulation of a single-station queue satisfying Assumption~\ref{asm:single}, and $\rho=\lambda/(m\mu)<1$. Suppose we are using backward or forward Euler approximation schemes to compute the long-run average number of customers in the system. Then the relative approximation error is bounded from above by  $\overline{\mathrm{RE}}(h)=\lambda h/E[N]$, which satisfies
\begin{equation}
\setlength{\abovedisplayskip}{3pt}
\setlength{\belowdisplayskip}{3pt}
\overline{\mathrm{RE}}(h)\leq \mu h.\label{eq:re_bound}
\end{equation}
Furthermore, assume $m\rightarrow\infty$, and $\rho\rightarrow 1$ from below.
\begin{enumerate}
\item[(i)] If $m(1-\rho)\rightarrow\infty$, then
\[
\frac{\overline{\mathrm{RE}}(h)}{\mu h}\rightarrow 1.
\]
\item[(ii)] If $m(1-\rho)\rightarrow\beta\in(0,\infty)$, then
\[
\frac{\overline{\mathrm{RE}}(h)}{\mu h}\rightarrow \frac{\beta}{1+\beta}.
\]
\item[(iii)] If $m(1-\rho)\rightarrow 0$, then
\[
\frac{\overline{\mathrm{RE}}(h)}{m(1-\rho)\mu h}\rightarrow 1.
\]
\end{enumerate}
\end{theorem}

Theorem~\ref{thm:relative_error} addresses the accuracy of the backward and forward Euler schemes when used to approximate the long-run average number of customers in terms of relative approximation errors. For fixed step size $h$, even though the absolute error bound $\lambda h$ (where $\lambda =\rho m \mu$) explodes as the arrival rate $\lambda$ goes to infinity with the scaling of the system, inequality~\eqref{eq:re_bound} indicates that the relative errors remain capped at $\mu h$. This result underscores the stability and reliability of using our Euler schemes to approximate performance measures in large queueing networks. The theorem ensures that, as the network scales, the relative error does not spiral out of control, thereby affirming the robustness of the Euler methods in large-scale applications.

The theorem then considers the asymptotic behavior of the relative error bound $\overline{\mathrm{RE}}(h)$ by assuming the number of servers $m$ goes to infinity while the system utilization $\rho$ approaches 1 from below. This setup suggests that the queueing system is expanding its capacity with an increasing number of servers, and the traffic intensity is growing correspondingly, yet never exceeds 1. This assumption places the system in a heavy traffic regime, a condition where the relative performance of the approximation schemes is particularly critical to evaluate as the system operates near its maximum capacity. It turns out that  $\overline{\mathrm{RE}}(h)$ exhibits different asymptotic behaviors along three sub-regimes, which are differentiated based on the relationship between the rates at which $m$ and $\rho$ increase.

\begin{itemize}
\item Regime (i): This corresponds to situations where the increase in the number of servers $m$ outpaces the increase in $\rho$ towards 1.  In this case, the theorem suggests that the bound on the relative error becomes asymptotically equivalent to $\mu h$, making the bound~\eqref{eq:re_bound} tight.

\item Regime (ii): This regime corresponds to the situations when the rate of increase in $m$ and the rate of increase in $\rho$ are balanced. In this case the relative error's bound is lower than what the general bound $\mu h$ would suggest. It still scales with the step size $h$, but the proportionality constant is smaller than $\mu$.

\item Regime (iii): In this regime, $\rho$ approaches 1 more rapidly than $m$ increases, and the result implies that $\overline{\mathrm{RE}}(h)/(\mu h)\rightarrow 0$. This suggests that in systems operating with high utilization, the upper bound on the approximation error becomes progressively tighter as system size increases. This enhanced accuracy in approximation arises because the expected number of customers in the system, $E[N]$, grows at a rate that outpaces the increase in the absolute approximation error. Consequently, the error relative to the scale of the system (i.e., relative to $E[N]$) becomes smaller, implying that the larger the system, the more precise the approximation becomes in a relative sense.
\end{itemize}

Note that the sub-regimes we consider differ from those defined in \cite{halfinwhitt1981}. In this seminal work, Halfin and Whitt identify three distinct regimes for a queueing system as it scales: (a) $\sqrt{m}(1-\rho) \rightarrow \infty$, (b) $\sqrt{m}(1-\rho) \rightarrow \beta' \in (0,\infty)$, and (c) $\sqrt{m}(1-\rho) \rightarrow 0$. They demonstrate that the asymptotic behavior of the probability $P(N > m)$ is uniquely characterized in each of these regimes. In contrast, our classification of regimes takes into account not only the scaling of $N$ but also the magnitude of absolute errors in Euler approximations. As $m \rightarrow \infty$, it is evident that regimes (a) and (b) from Halfin and Whitt's framework fall into our regime (i), while regime (c) intersects with all three regimes in our analysis. Specifically, under the ``square root staffing rule" corresponding to Halfin and Whitt's regime (b) and our regime (i), the probability of delay stabilizes to a nondegenerate limit, whereas the relative error bound \eqref{eq:re_bound} becomes asymptotically tight.

\subsubsection{Queueing Network}
The next theorem extends the relative error analysis to the simulation of queueing networks. Its proof can be found in Appendix~\ref{proof:relative_error_network}.

\begin{theorem}\label{thm:relative_error_network}
Consider the simulation of a queueing network satisfying Assumption~\ref{asm:network}, and $\rho_i=\tilde\lambda_i/(m_i\mu_i)<1$ for $i=1,\cdots,n$. Suppose we use backward or forward Euler approximation schemes to compute the long-run average number of customers in the system. Then the relative approximation error is bounded from above by  $\overline{\mathrm{RE}}(h)=\sum_{i=1}^n\tilde\lambda_i h/E[N]$, which satisfies
\begin{equation}\label{eq:re_bound_network}
\overline{\mathrm{RE}}(h)\leq \frac{\sum_{i=1}^n m_i\rho_i\mu_i }{\sum_{i=1}^nm_i\rho_i}h\leq \left(\max_{1\leq i\leq n}\mu_i\right) h.
\end{equation}
\end{theorem}

Theorem~\ref{thm:relative_error_network} confirms that the relative error of the backward or forward Euler schemes, when used to approximate the long-run average number of customers in a feedforward queueing network, also decays in proportion to the step size $h$. The proportionality constant can be expressed as the weighted average service rate of individual servers in the network, and thus bounded by the server with the highest service rate. Regarding to the asymptotic behavior of the relative error bound, notice that
\[
\setlength{\abovedisplayskip}{3pt}
\setlength{\belowdisplayskip}{3pt}
\min_{i=1,\cdots,n}\overline{\mathrm{RE}}_i(h)\leq \overline{\mathrm{RE}}(h)\leq \max_{i=1,\cdots, n}\overline{\mathrm{RE}}_i(h),
\]
where $\overline{\mathrm{RE}}_i(h)=\tilde \lambda_i h/E[N(i)]$ represents the relative error bound for the $i$th node. Therefore, if all nodes in the network operate within the same asymptotic regime, the results derived for a single-station queue in Theorem~\ref{thm:relative_error} can be applied to the entire network. Otherwise,  if different nodes fall into different asymptotic regimes, the overall asymptotic behavior of $\overline{\mathrm{RE}}(h)$ can be dominated by the nodes where the number of customers increases most rapidly. In particular, if these critical nodes operate under regime (iii), where the relative error decays at a rate faster than $\mu_i h$, the network's overall relative error bound may similarly demonstrate a rapid decay.

\subsection{Selection Rule for Time Step}\label{sec:selection_rule}
Theorem~\ref{thm:relative_error_network} offers a practical guideline for choosing the step size $h$ in Euler approximation methods for large queueing networks. A straightforward strategy for selecting $h$ can be formalized as follows:

\vspace{10pt}
\textbf{Selection Rule for $h$}: Given the maximum service rate across all servers, denoted by $\bar{\mu}$, and a predefined target for the relative error, denoted by $\alpha_m$, the step size $h$ should be set as $h=\alpha_m/\bar \mu$. 
\vspace{10pt}

This rule is designed to ensure that the relative errors in the Euler schemes, when estimating the long-run average number of customers, are controlled within the specified tolerance level. Notice that the subscript $m$ in $\alpha_m$ emphases that the tolerance level can be dependent on the system size. For instance, setting $\alpha_m={\rm O}(1/\sqrt{m})$ ensures that the relative error converges to zero as $m$ goes to infinity. Nonetheless, this rule may be conservative, as the actual relative error could decay more rapidly than predicted by \eqref{eq:re_bound_network}.
In practice, it may be advantageous to adopt a larger value of $h$ to improve computational efficiency. To optimize the trade-off between accuracy and efficiency, a sensitivity analysis can be conducted. This process involves incrementally adjusting $h$ and observing the effects on the simulation outcomes. Such approach allows for the fine-tuning of $h$ beyond the conservative estimate provided by the initial rule, potentially leading to faster simulations that still meet the desired accuracy requirements.

\section{Simulation of Sojourn Time with Euler Approximation}\label{sec:sojourn}
In our approach, we aggregate the arrival and departure events to reduce the computational effort. Consequently, detailed individual-level records, such as the total sojourn time of a given customer, are not maintained in the simulated queueing network. However, we can still generate samples of the sojourn time for individual customers who arrive at a certain node during a certain time interval. To illustrate the practical importance of our sojourn time simulation approach, consider the following scenario. Imagine we have employed the algorithm in Subsection~\ref{subsec:sim_net} and generated 1,000 sample paths of system state for a queueing network over the next day. As the decision makers, we want to assess the probability of customers experiencing lengthy sojourn time if they arrive between 13:05 and 13:10 on that day. Based the existing simulation data on system state, our approach can generate samples of sojourn time for such customers with minimal additional cost. 

In general, suppose we have simulated several sample paths of system state $\boldsymbol{N}_\tau$ and recorded the aggregated number of departures $\boldsymbol{D}_\tau$ for a feedforward queueing network. 
Let's consider a customer enters the system through node $i_0$ during the time interval $((\tau_0-1)h,\tau_0 h]$. We aim to generate samples of this customer's sojourn time using the previously mentioned sample paths. The sampling procedure is given by Algorithm~\ref{algo:sojourn}.

Basically, the above procedure samples this customer's position at each time step, based on the conditional probability derived in Lemma~\ref{lem:pure-departure-3}.
Similar to the algorithms in Section~\ref{sec:sim_state} for simulating the number of customers, this procedure for simulating sojourn time is also a time-discretization method, and hence the computational cost is proportional to the number of time intervals. The computational complexity is therefore no more than the simulation algorithms for system state. Furthermore, by making use of the exiting sample path data on system state, the additional cost for simulating sojourn time is minimal, further underscoring the effectiveness of our simulation framework.

\begin{algorithm}[H]
\caption{Euler Approximation of Sojourn Time}\label{algo:sojourn}
\SetKwBlock{WaitingStep}{Waiting Step:}{end}
\SetKwBlock{ServiceStep}{Service Step:}{end}
\SetKwBlock{SetupStep}{Setup Step:}{end}
{\bf Initialization:} Set sojourn time $T=0$. Let $i=i_0$ and $\tau = \tau_0$\;
\BlankLine
\SetupStep{
Read $N_{\tau,i}$\;
\eIf{$N_{\tau,i}> m_i$}{
Go to Waiting Step;
}{
Go to Service Step;
}
}
\BlankLine
\WaitingStep{
  Compute $\eta = \min\{j\geq1: N_{\tau,i}-D_{\tau+1,i}-\cdots-D_{\tau+j,i}\leq m_i\}$\;
    $T\gets T+\eta h$\, $\tau\gets \tau+\eta$, and update $N_{\tau,i}$\;
    Go to Service Step\;}
\BlankLine
\ServiceStep{
  Generate $I\sim \mathrm{Bernoulli}(p_{stay})$, where
  \[
    p_{stay} =
    \begin{cases}
    \dfrac{N_{\tau,i} - D_{\tau,i}}{N_{\tau,i}}, & \text{if }N_{\tau,i}\leq m_i\\
    \left(\dfrac{m_i - 1}{m_i}\right)^{N_{\tau,i}-m_i}\dfrac{N_{\tau,i} - D_{\tau,i}}{m_i},&\text{if }m_i<N_{\tau,i}\leq m_i+ D_{\tau,i}\\
    \left(\dfrac{m_i - 1}{m_i}\right)^{D_{\tau,i}}, & \text{if }N_{\tau,i}> m_i+ D_{\tau,i}\\
    \end{cases}\;
  \]
  \eIf{$I=1$}{
    $T\gets T+ h$\, $\tau\gets \tau+1$\;
    Go to the beginning of Service Step;
  }{
    Set $i' = j$ with probability $R_{\tau,i,j}/D_{\tau,i}$\;
    $T\gets T+ h$, $\tau\gets \tau+1$\;
    \eIf{$i'$ refers to another node}
    {
    $i\gets i'$\;
      Go to Setup Step;
    }
    {Return $T$ as the sojourn time and terminate the algorithm;
    }
  }
}
\end{algorithm}
\vspace{3pt}

\section{Numerical Experiments}\label{sec:exp}
In this section, we assess the efficacy of our proposed algorithms from diverse perspectives. The simulation programs are developed using the Python programming language, and the experiments are conducted on a Linux-based computer system with 64 CPU cores and 512 GB of RAM.

\subsection{Run-Time and Approximation Error under Different $h$}\label{subsec:run-time_interval-result}
We initiate our experiments by testing backward and forward approximation schemes at various $h$ values on a network characterized by $n=1,000$ and $m=200$. The run-time and relative error results are documented in Table \ref{tab:h-time-error}. Simulations encompass a time horizon of 1,000 units, and results for both schemes are averaged over 20 independent replications.

\begin{table}[htb]
\caption{Run-Time and Approximation Error under Different $h$}\label{tab:h-time-error}
\centering
\begin{tabular}{ccccccc}
\toprule
\multirow{2}{*}{Time Interval} & \multicolumn{2}{c}{Backward} & \multicolumn{2}{c}{Forward} & \multicolumn{2}{c}{Average}                   \\
\cmidrule(l){2-3} \cmidrule(l){4-5} \cmidrule(l){6-7}
                               & Run-Time   & Relative Error  & Run-Time  & Relative Error  & Run-Time & Relative Error\\
\midrule
0.2 &  8.6s & 14.12\% & 20.8s & -12.28\% & 21.2s & 0.92\% \\
0.1 & 18.1s & 6.81\%  & 42.3s & -6.39\%  & 43.0s & 0.21\% \\
0.05& 33.8s & 3.35\%  & 82.0s & -3.25\%  & 83.4s & 0.051\% \\
0.02& 80.8s & 1.32\%  & 199.4s & -1.32\% & 202.4s& 0.004\% \\
0.01& 158.1s& 0.66\%  & 380.2s & -0.66\% & 386.2s& -0.001\% \\                
\bottomrule
\end{tabular}
\end{table}

In both the forward and backward schemes, we observe that the run-time exhibits an inverse proportionality to the time interval, aligning with the computational complexity outlined in Subsection~\ref{subsec:complexity}. Additionally, the relative error decreases linearly as the interval diminishes, in accordance with theoretical analysis. As discussed in Section \ref{sec:simu_vec}, despite equal complexities in both schemes, the backward scheme demonstrates significantly shorter run-time. This discrepancy arises from the layer-by-layer progression inherent in the forward approximation, which impedes efficient utilization of vectorization benefits.

Additionally, we conduct experiments by simultaneously employing forward and backward schemes on 2 processes. Subsequently, we calculate the average customer counts obtained from both schemes. We refer to this approach as the \emph{average scheme}. The corresponding results are also presented in Table \ref{tab:h-time-error}. The average scheme exhibits only a marginal increase in run-time compared to running the forward scheme alone. However, it effectively mitigates the relative approximation error. In practical applications, employing the average scheme allows for the utilization of a larger time interval, thereby decreasing the run-time while maintaining the same target error level.

\subsection{Run-Time Comparison with DES}\label{subsec:run-time-result}
We then conduct a run-time comparison between our approximation approach and Ciw, which stands as one of the state-of-the-art open-source tools for simulating queueing networks \citep{palmer2019ciw}. 
We employ the average scheme in our approximation approach. Furthermore, we use a uniform time interval of {0.15} across all network scales. 
The simulations are conducted over a time horizon of {100} time units. And the run-time results of both methods under varying network parameters, averaged over 20 independent replications, are showcased in Table \ref{tab:run-time-comp}. In certain scenarios, the run-time of DES is exceedingly long, sometimes surpassing a duration of 24 hours. In these instances, we halt the simulation, and the associated run-time is documented as ``$>24$hr". Approximation error information is omitted from the table, and all observed relative errors are below 1\%.

\begin{table}[htb]
\caption{Run-time comparison with DES}\label{tab:run-time-comp}
\centering
\begin{tabular}{cccccccc}
\toprule
\multirow{2}{*}{Parameters}     
                                & $n$ & \multicolumn{3}{c}{100} & \multicolumn{3}{c}{1,000} \\
\cmidrule(l){2-2} \cmidrule(l){3-5} \cmidrule(l){6-8}
                                & $m$ & \multicolumn{1}{c}{20} & \multicolumn{1}{c}{200} & \multicolumn{1}{c}{1,000} & \multicolumn{1}{c}{20} & \multicolumn{1}{c}{200} & \multicolumn{1}{c}{1,000} \\
\midrule
\multicolumn{2}{c}{Euler Approximation} & $1.2$s & $1.2$s & $1.2$s & $2.6$s & $2.9$s & $2.9$s \\
\multicolumn{2}{c}{DES}             & $33.5$s & $5,911.3$s & $>24$hr & $4,378.0$s & $>24$hr & $>24$hr \\
\multicolumn{2}{c}{Speedup}             & $27.9$ & $4,926.1$ & $>72,000$ & $1,683.8$ & $>30,000$ & $>30,000$ \\
\bottomrule
\end{tabular}
\end{table}

When the number of servers increases for the same number of nodes, the run-time of the Euler approximation remains relatively stable, given the consistent adoption of the same time interval. Contrary to the expectations based on complexity analysis, a tenfold increase in the number of nodes does not result in a tenfold increase in run-time of the Euler approximation; instead, it only rises by less than threefold. This discrepancy can be attributed to the more significant acceleration achieved through vectorization when dealing with larger vector sizes. Notably, traditional DES proves impractical for large-scale networks due to its slow speed. In contrast, our proposed Euler approximation approach exhibits significantly enhanced simulation speed when compared to DES in large-scale networks, with relatively minor approximation errors.

\subsection{Relation between $h$ and Approximation Error}
\begin{figure}[htb]
{
\centering
\includegraphics[scale=0.37]{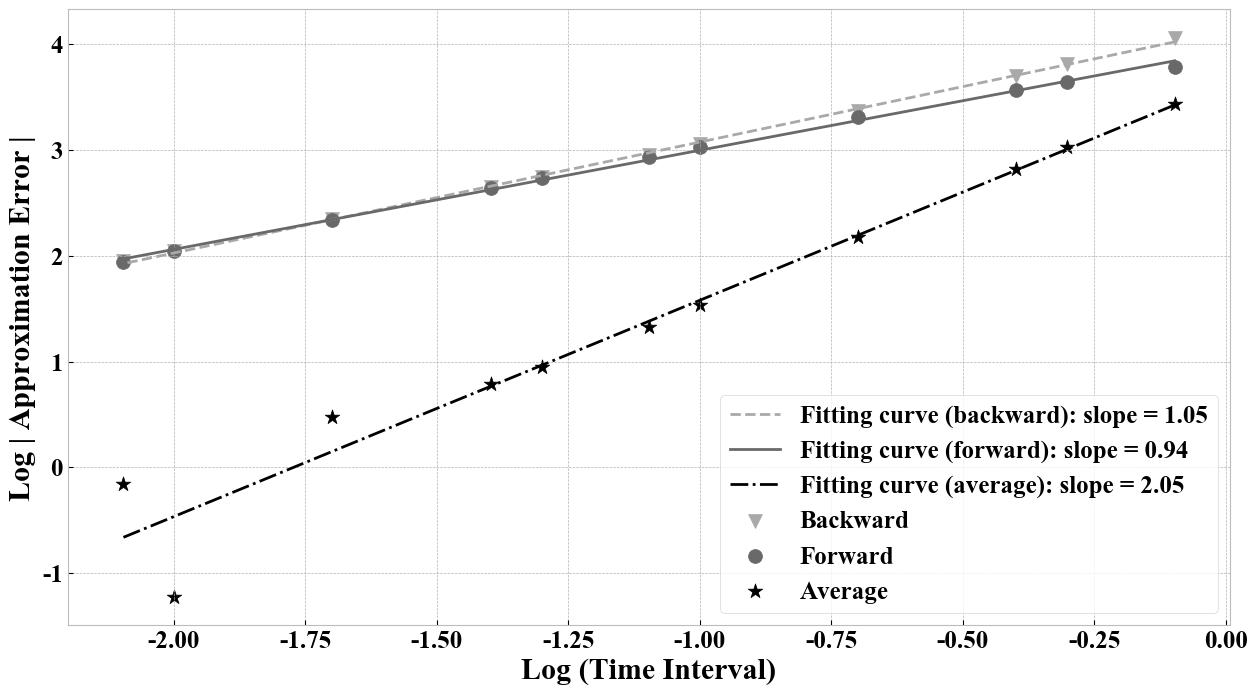}
\caption{Relation between $h$ and approximation error.\label{fig:error-h}}
}
\end{figure}

To further investigate the relationship between the approximation error and the time interval, we conduct experiments using backward, forward and average approximation schemes with various $h$ values on a network characterized by $n=100$ and $m=200$. The simulations span a time horizon of 1,000 units, and the approximation error results for the three schemes are averaged over 20 independent replications. We then generate a log-log scatter plot depicting the variation of the approximation error with respect to the time interval. Additionally, a linear fit is applied to the data points corresponding to each scheme, and the slope information is presented in Figure \ref{fig:error-h}.

The slopes of the fitting curves for both backward and forward approximation schemes closely approximate 1. This suggests that the approximation errors of the two schemes diminish linearly with the step size $h$, aligning with the characteristics of the conventional Euler scheme employed in simulating diffusion processes. An interesting discovery emerges when averaging the results from both backward and forward schemes: the rate at which the approximation error diminishes over the time interval can achieve a second-order magnitude. The theoretical underpinning of this assertion is an interesting topic for future research.

\subsection{Convergence of Relative Error}\label{subsec:relative-error-result}

\begin{figure}[htb]
{
\centering
\includegraphics[scale=0.35]{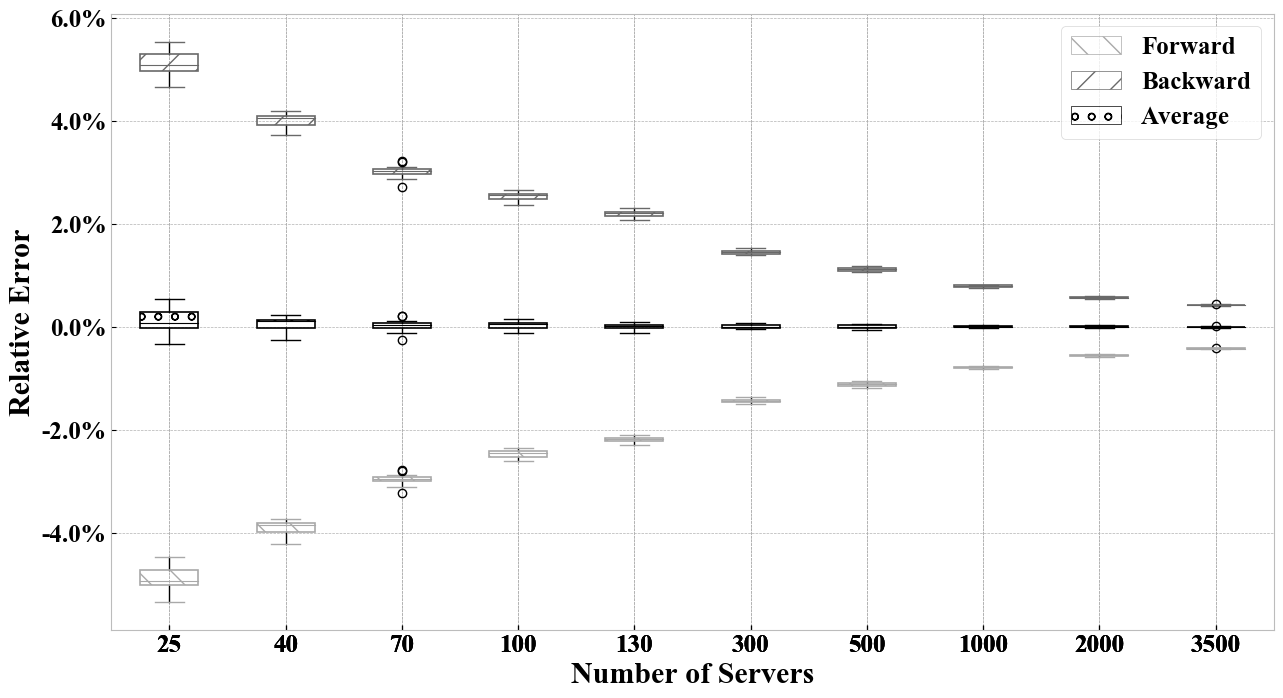}
\caption{Convergence of Relative errors with Decaying Time Intervals.\label{fig:relative_error}}
}
\end{figure}

In this subsection we assess the relative approximation error of our proposed approximation algorithms on a single-station queue 
with varying numbers of servers ($m$) per node. The simulations span a time horizon of 1,000 units. Time intervals are configured according to recommended selection rule, i.e., $h=\alpha_m/\bar \mu$, where $\alpha_m=0.4/\sqrt{m}$. We calculate the relative errors between the simulated network's average total customer count and the theoretical value, and boxplots of 20 independent replications are illustrated in Figure \ref{fig:relative_error}. Notably, the relative error consistently decreases toward zero with an increasing number of servers, aligning with our theoretical analysis. 

In addition, we take the average of customer counts obtained from both forward and backward approximations. We then depict the relative error between the obtained results and the theoretical values as the number of servers varies, as illustrated in Figure \ref{fig:relative_error}. The graph reveals that the result, after averaging, exhibits significantly reduced relative error when contrasted with both the forward and backward approximations. Furthermore, despite the asymptotic analysis indicating that the relative error approaches zero only as the number of servers approaches infinity, the averaged result attains a remarkably low relative error even with a relatively modest number of servers.

\subsection{Distribution of Number of Customers }\label{subsec:len-result}
\begin{figure}[htb]
{
\centering
\includegraphics[scale=0.5]{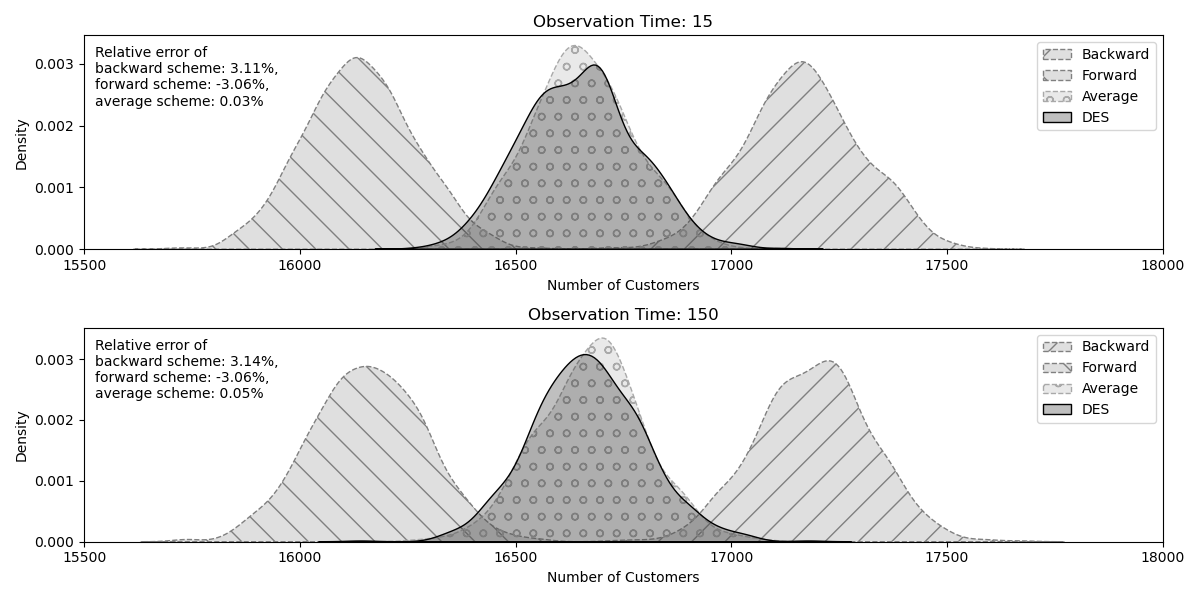}
\caption{Number of Customers Performance of Euler Approximation.\label{fig:len_dist}}
}
\end{figure}

To further evaluate the precision of our approximation algorithms, we perform 1,000 repeated experiments on a network with $n=100$ and $m=200$. These experiments utilize both our fast approximation method and DES over a time span of 200 units. Subsequently, we create kernel density estimate (KDE) plots depicting the customer counts within the network at two time points: $t=15$ and $t=150$. Figure \ref{fig:len_dist} depicts the performance of our method in the three approximation schemes.

The figure illustrates that, in the backward scheme, our method produces a customer count distribution closely resembling that of DES, albeit with a rightward shift. Similarly, in the forward scheme, the distribution closely resembles DES, albeit with a leftward shift. Upon averaging, our method's results closely align with those of DES.

Furthermore, across all three schemes, variations in the observation time have a negligible impact on the performance of the customer count distribution. This indicates that our method's approximation error does not accumulate over time.

\subsection{Distribution of Sojourn Time }\label{subsec:sojourn-result}
\begin{figure}[htb]
{
\centering
\includegraphics[scale=0.54]{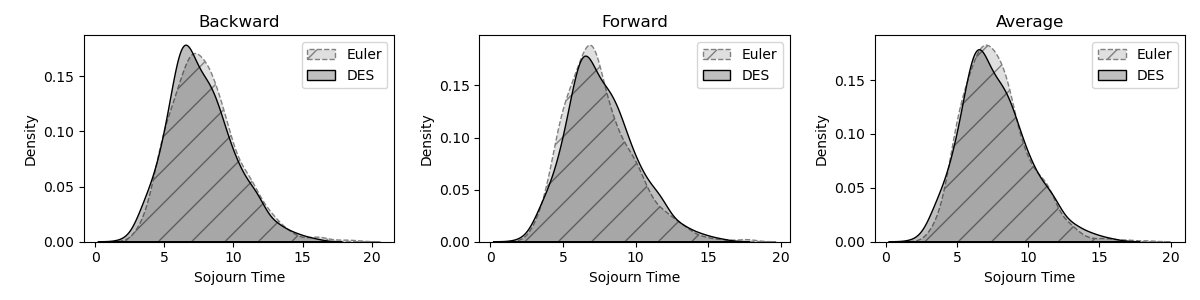}
\caption{Sojourn Time Performance of Euler Approximation.\label{fig:sojourn}}
}
\end{figure}

Our method is also capable of simulating customer sojourn times. To evaluate the performance, we conduct 1,000 repeated experiments on a network with $n=100$ and $m=200$ using both our fast approximation method and DES over a time span of 200 units. Using the sample paths of number of customers and departure counts generated by our method, we compute the sojourn time for a customer arriving at node 1 at observation time $t=150$ following the procedure outlined in Section \ref{sec:sojourn}. We generate KDE plots illustrating the customer's sojourn time using both DES and our proposed method. The figures illustrating the backward, forward, and average schemes are depicted in Figure \ref{fig:sojourn}. As shown in the figures, the approximation performance of the sojourn time distribution in all three schemes is evidently satisfactory.

\vspace{-5pt}
\vspace{-10pt}

\subsection{Performance of the Algorithms on Real-World Queueing Networks}\label{subsec:real-net}
In the preceding experiments, the networks are generated  with an equal number of servers per node to test mainly the theoretical properties. In this subsection we will evaluate the performance of our method on real-world networks by examining two practical network examples.

\subsubsection{A Healthcare Network}
\begin{figure}[htb]
{
\centering
\includegraphics[scale=0.38]{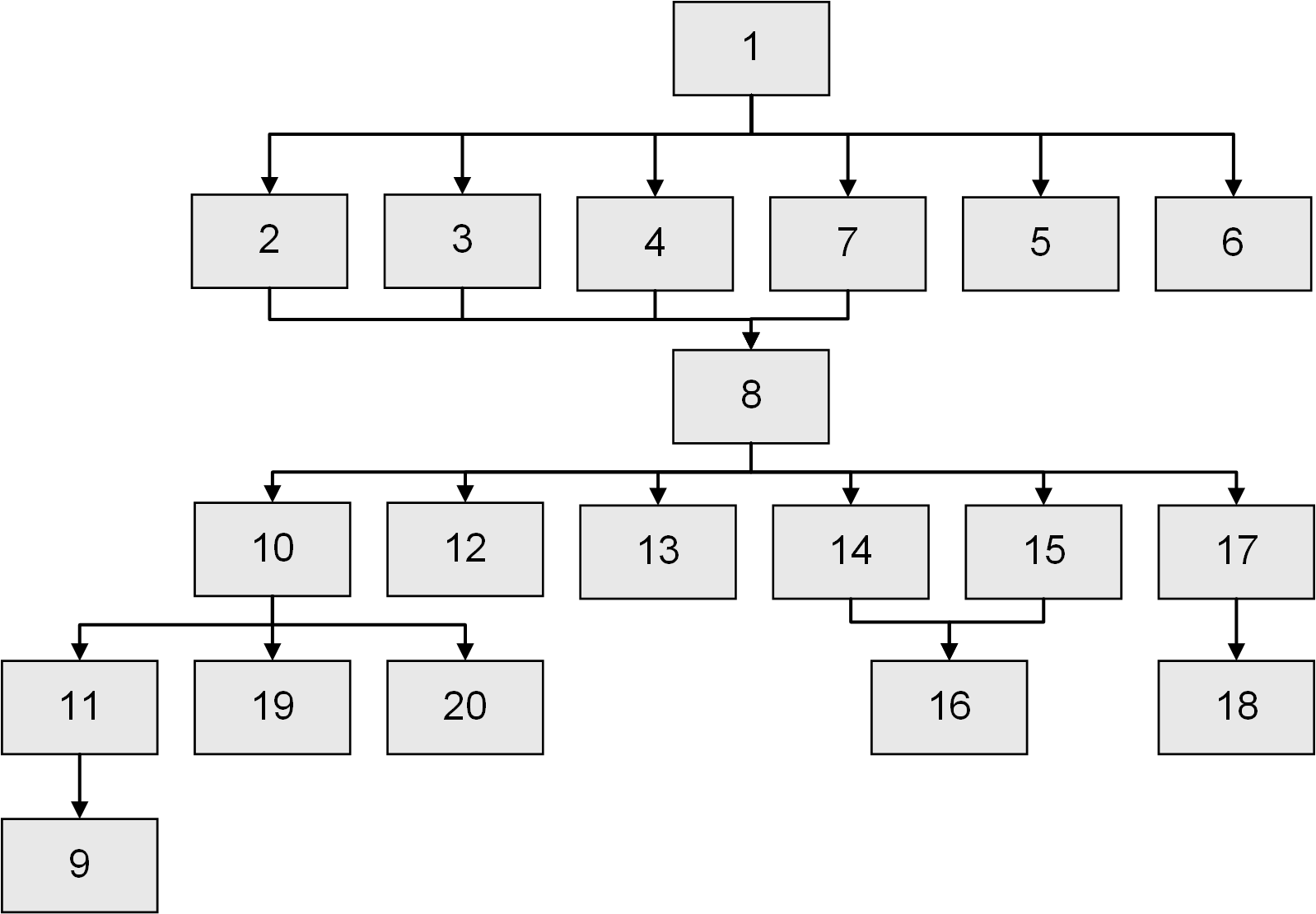}
\caption{A Hospital Operation Network.\label{fig:hospital}}
}
\end{figure}

In a hospital, the operational dynamics of patient flow can be typically represented as a queueing network. Here, medical units serve as network nodes, patients as customers, and beds, medical staff, and medical equipment as servers \citep{armony2015patient}. We use the hospital described in \cite{alenany2017modelling} for our testing, and it comprises 20 departments. We have made slight adjustments to transform the operation network into a feed-forward queueing network, as illustrated in Figure \ref{fig:hospital}. Detailed node information for the hospital's operation network is provided in Table \ref{tab:hospital-info}.

\begin{table}[htb]
\footnotesize
\caption{Node Information for The Hospital Operation Network}\label{tab:hospital-info}
\centering
\renewcommand{\arraystretch}{1.1}
\begin{tabular}{m{1.5cm}<{\centering}m{3.5cm}<{\centering}m{2cm}<{\centering}m{1.5cm}<{\centering}m{3.7cm}<{\centering}m{2cm}<{\centering}}
\toprule
Node & Station & Number of Servers ($m$) & Node & Station & Number of Servers ($m$) \\
\midrule
1    & Triage                      & 10   &
11   & Operation Room              & 100   \\
2    & Internal Medicine Room      & 4   &
12   & Fixation Room               & 6   \\
3    & Surgery Room                & 50   &
13   & Internal Department         & 200 \\
4    & Ophthalmology Room          & 3   &
14   & Intermediate Burn Care Unit & 12  \\
5    & Ear/Nose/Throat Room  & 3   &
15   & Intensive Burn Care Unit    & 4   \\
6    & Orthopedics Room            & 3   &
16   & Burn OR                     & 3   \\
7    & Resuscitation Room          & 20   &
17   & Orthopedic Care Unit        & 6   \\
8    & Management Rooms            & 90   &
18   & Orthopedic OR               & 12   \\
9    & I.C.U.                      & 200  &
19   & Ophthalmology OR            & 12   \\
10   & Pre/Post-Operative Care     & 120  &
20   & Ear/Nose/Throat OR                      & 3  \\
\bottomrule
\end{tabular}
\end{table}

We conduct simulations using both DES and our proposed method, over a time horizon of 1,000 time units. For our method, the average 
scheme is adopted and the time interval is set to {0.2}. 
Run-time results are recorded, and relative errors are calculated by comparing the average total number of customers in the simulated network to the theoretical value. Both results, averaged from 20 independent replications, are presented in Table \ref{tab:run-time-real-net}. The Euler approximation demonstrates a run-time approximately one-tenth that of DES, accompanied by a correspondingly negligible relative error in the average total number of customers.

\subsubsection{A Data Center Network}
Currently, over 5 billion users rely on uninterrupted Internet connectivity. Data centers have become indispensable and critical infrastructure assets that drive the continuous growth of Internet services and applications \citep{xia2016survey}. Consequently, the analysis of data center networks has gained immense significance. 

Traditional data center networks are typically constructed based on hierarchical topology with three switch layers \citep{nooruzzaman2021hyperscale}: access switches, which link to the servers; aggregation switches, also known as distribution routers, connecting to the access switches; and core switches interconnecting the aggregation switches. With the expansion of data center sizes, fabric architecture has gained popularity in modern data centers. 

\begin{figure}[htb]
{
\centering
\includegraphics[scale=0.28]{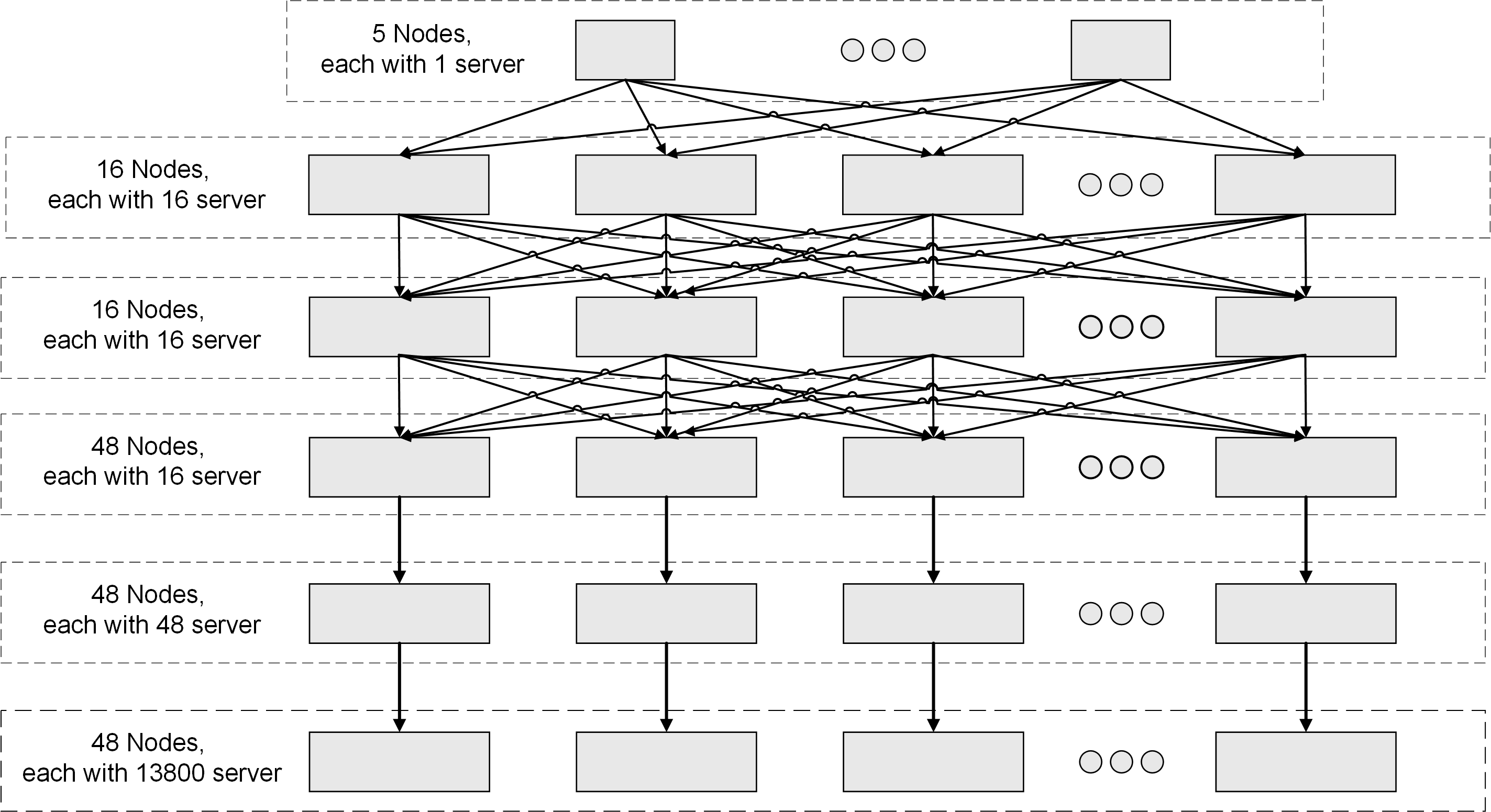}
\caption{A Data Center Network.\label{fig:data-center}}
}
\end{figure}

In this context, we select the Facebook F16 Fabric Aggregator Architecture \citep{nooruzzaman2021hyperscale} and abstract it into the queueing network illustrated in Figure \ref{fig:data-center}, for simulation tests. We conduct simulations using both DES and our proposed method, over a time horizon of 100 time units. For our method, the average scheme is adopted and the time interval is set to {0.05}.
Speedups are derived from run-time results averaged across 20 independent replications, and relative errors are calculated by comparing the average total number of customers in the simulated network to the theoretical value. These results are presented in Table \ref{tab:run-time-real-net}. The scale of the network surpasses the capability of traditional DES. As the simulation run-time surpasses 24 hours, we terminate the process. The speedup result presented in the table is computed with a reference run-time of 24 hours. Simultaneously, our Euler approximation approach exhibits minimal run-time requirements, and the approximation error is negligible, thereby reinforcing the superiority of our approach.

\begin{table}[htb]
\caption{Results for Real-World Queueing Networks}\label{tab:run-time-real-net}
\centering
\begin{tabular}{ccc}
\toprule
\multirow{3}{*}{Parameters}   & Hospital Network  & Data Center Network   \\
 \cmidrule(l){2-2} \cmidrule(l){3-3} 
 & $n=20, m_i^{max}=200$  & $n=181, m_i^{max}=13,800$   \\
  \cmidrule(l){2-2} \cmidrule(l){3-3} 
  & Time Horizon: 1,000 units  & Time Horizon: 100 units  \\
\midrule
Run-Time of Euler    & 4.4s & 2.6s       \\
Run-Time of DES    & 41.7s & $>24$hr    \\
Speedup & 9.5 & $>33,230$ \\
Relative Error & 0.89\% &  -0.84\% \\
\bottomrule
\end{tabular}

\end{table}

\section{Concluding Remarks}\label{sec:conclude}

Motivated by the near-universal applicability of Euler approximation in simulating continuous systems, this paper investigates the potential of applying Euler approximation to simulate discrete systems, with a particular emphasis on Markovian queueing networks. We develop two simulation schemes: the backward scheme and the forward scheme. By the aggregation of time intervals, these two schemes achieve a significant reduction in computational complexity for simulating large systems; by decoupling of event interactions, the efficiency of these schemes can be further enhanced by vectorization.
Theoretical analysis confirms that our discrete-system adaptations of Euler approximation are first-order methods, with approximation errors decay in proportion to the step size. Interestingly, we find that the absolute approximation errors maintain uniform bound over time. Moreover, with the recommended choice of time step, 
the asymptotic relative error of our simulation schemes
approaches zero as the network scales up, while maintaining a much lower computational complexity
compared to the traditional DES. Therefore, our methods hold particular value in managing large-scale and complex queueing networks, especially when the primary focus is to comprehend the overall system performance in a timely manner.

Our research opens the door to several promising directions for future study. Firstly, while the current paper is focused on Markovian systems, the potential for adapting our method to non-Markovian systems remains an interesting prospect that calls for more in-depth investigation. Secondly, insights gained from our numerical experiments point toward the possibility of attaining second-order approximation by averaging the forward and backward estimators. This observation lays the groundwork for extending our theoretical framework and algorithm design to formulate methods of higher-order approximation.

\bibliographystyle{informs2014} 
\bibliography{main} 

\newpage

\begin{APPENDICES}
\setcounter{page}{1}

\section{Vectorized Version of GenerateDeparture (Algorithm \ref{alg:pure-departure}) }\label{sec:pure_departure_vec}
In this section we aim to provide a vectorized version of Algorithm~\ref{alg:pure-departure}, which takes the vector-valued inputs $\boldsymbol{x}$, $\boldsymbol{\mu}$, and $\boldsymbol{m}$ with the same size, and the scalar input $h>0$, and generates a vector output $\boldsymbol{D}$. We assume the components of $\boldsymbol x$, $\boldsymbol{\mu}$, and $\boldsymbol{m}$ are non-negative integers, positive real numbers, and positive integers, respectively.
The vectorized algorithm proceeds as follows:
\begin{align*}
\boldsymbol{T} &= \mathrm{Erlang}\left( ({\boldsymbol{x} - \boldsymbol{m})^+, \boldsymbol{m} \times \boldsymbol{\mu} } \right) \nonumber;\\
\boldsymbol{D} & = \left[\mathrm{Binomial}\left(\min\{\boldsymbol{x},\boldsymbol{m}\}, 1-\exp\{-\boldsymbol{\mu}\times(h-\boldsymbol{T})\}\right)+(\boldsymbol{x} - \boldsymbol{m})^+\right]\times {1_{\left\{ {\boldsymbol{T} \le h} \right\}}}\\
&~~~~+\mathrm{Binomial}\left( (\boldsymbol{x}-\boldsymbol{m}-1)^+, h/\boldsymbol{T} \right) \times {1_{\left\{ {\boldsymbol{T} > h} \right\}}}.
\end{align*}
In this context, all operations between scalars and vectors or between vectors are to be interpreted as element-wise. The functions $\mathrm{Erlang}$ and $\mathrm{Binomial}$ are capable of generating multivariate random variables through vectorized operations, consistent with the functionality provided by many computational software packages, such as MATLAB and the Python library NumPy. By convention, these functions return zero if the first argument is zero. Note that $1_{\{\cdot\}}$ represents a vectorized indicator function, which outputs a vector of 0s and 1s depending on whether the condition inside is true for the corresponding element, ensuring that the operation is applied element-wise across vectors. It is straightforward to check that each component $D_i$ of $\boldsymbol{D}$ satisfies:
\[
D_i \sim \mathrm{GenerateDeparture}(x_i,m_i,\mu_i,h).
\]

\section{Proof of Lemma~\ref{lem:pure-departure-1}}\label{proof:pure-departure-1}
\proof{Proof of Lemma~\ref{lem:pure-departure-1}.}
(i) If $x\leq m$, then initially the $x$ customers are at service at the beginning of the time period. Furthermore, the probability that a customer will complete service in a time period of $h$ is $1-e^{-\mu h}$. Therefore, the number of customers completing service in a time period of $h$ follows binomial distribution $\mathrm{Binomial}(x, 1-e^{-\mu h})$.

(ii) If $x\geq m+1$, then initially $m$ customers are at service and $x-m$ customers are waiting in the queue. In addition, all servers are busy up to time $T$, when the $(x-m)$th departing customer leaves the system, with $m$ customers at service and no one in the queue.
Therefore, the customers leaves the system according to a Poisson process with rate $m\mu$ from time 0 to time $T$, and given that the $(x-m)$th departing customer leaves the system at time $T$, the departing times of the first $(x-m-1)$ customers have the same distribution as $(x-m-1)$ uniform random variable over $(0,T)$.

If $T>h$, then each of the first $(x-m-1)$ customers leaves the system before $h$ with a probability of $h/T$. Therefore, the number of customers leaves the system before $h$ follows binomial distribution $\mathrm{Binomial}(x-m-1,h/T)$.

If $T\leq h$, then $x-m$ customers leave the system from time 0 to time $T$. Using the similar argument as the one in the proof of part (2), we know the number of customers leave the system from time $T$ to time $h$ follows  $\mathrm{Binomial}(m, 1-e^{-\mu (h-T)})$.
\Halmos
\endproof

\section{Proof of Lemma~\ref{lem:pure-departure-2}}\label{proof:pure-departure-2}
\proof{Proof of Lemma~\ref{lem:pure-departure-2}.}
\begin{enumerate}
\item[(i)] By conditioning on $X$, we have
\begin{align*}
P(\mathcal G_t(X)\geq x) &= \sum_{i=0}^\infty P(\mathcal G_t(i)\geq x)P(X=i)\\
&=\sum_{i=0}^\infty P(\mathcal G_t(i)\geq x)\left[P(X\geq i)-P(X\geq i+1)\right]\\
&=1+\sum_{i=1}^\infty \left[P(\mathcal G_t(i)\geq x)-P(\mathcal G_t(i-1)\geq x)\right]P(X\geq i).
\end{align*}
Since $X\preceq Y$, $P(X\geq i)\leq P(Y\geq i)$. In addition,  due to Proposition 12.6 in \cite{ross2019book}, we know a pure departure process increases stochastically in its initial state, so $P(\mathcal G_t(i)\geq x)-P(\mathcal G_t(i-1)\geq x)\geq 0$. Therefore
\begin{align*}
P(\mathcal G_t(X)\geq x) 
&\leq 1+\sum_{i=1}^\infty \left[P(\mathcal G_t(i)\geq x)-P(\mathcal G_t(i-1)\geq x)\right]P(Y\geq i)\\
&= \sum_{i=0}^\infty P(\mathcal G_t(i)\geq x)P(Y=i)\\
&=P(\mathcal G_t(Y)\geq x).
\end{align*}

\item[(ii)]  It suffices to prove that given $X=x\geq0$, we have
$$\mathcal G_{t+s}(x+1)\preceq \mathcal G_t(\mathcal G_s(x)+1)\preceq \mathcal G_{t+s}(x)+1.$$
Let $p_{x,y}(t)=P(\mathcal G_t(x)=y)$ be the transition probability function of the pure departure process. Then by Kolmogorov’s forward and backward equations, we have
\begin{align}
p'_{x,y}(t) &= q_{y+1,y}p_{x,y+1}(t)-q_{y,y-1}p_{x,y}(t),\label{eq:forward} \\
p'_{x,y}(t) &= q_{x,x-1}p_{x-1,y}(t)-q_{x,x-1}p_{x,y}(t),\label{eq:backward}
\end{align}
for any integers $x,y\geq 0$, where $q_{x,x-1}=\min\{x,m\}\mu$.
Given $h>0$ and integer $j\geq1$, define $F(s)=P(\mathcal G_{h-s}(\mathcal G_s(x)+1)\geq j)$ for $0\leq s\leq h$.  Then
\begin{align*}
F(s) &= \sum_{y=0}^x\sum_{z=j}^{y+1} P(G_s(x)=y) P(G_{h-s}(y+1)=z)= \sum_{y=0}^x\sum_{z=j}^{y+1} p_{x,y}(s) p_{y+1,z}(h-s).
\end{align*}
Taking derivative with respect to $s$, we get
\begin{align}
F'(s) &= \sum_{y=0}^x\sum_{z=j}^{y+1} \left\{p'_{x,y}(s) p_{y+1,z}(h-s)- p_{x,y}(s) p'_{y+1,z}(h-s)\right\}\notag\\
&=\sum_{y=0}^x\sum_{z=j}^{y+1} \left\{[q_{y+1,y}p_{x,y+1}(s)-q_{y,y-1}p_{x,y}(s)] p_{y+1,z}(h-s)\right.\notag\\
&~~~~~~~~\left.-p_{x,y}(s) [q_{y+1,y}p_{y,z}(h-s)-q_{y+1,y}p_{y+1,z}(h-s)]\right\}\notag\\
&=\sum_{y=0}^x\sum_{z=j}^{y+1} \left\{q_{y+1,y}p_{x,y+1}(s)p_{y+1,z}(h-s)- q_{y+1,y}p_{x,y}(s)p_{y,z}(h-s)\right\}\notag\\
&~~~~~~~~+\sum_{y=0}^x\sum_{z=j}^{y+1}[q_{y+1,y}-q_{y,y-1}]p_{x,y}(s)p_{y+1,z}(h-s), \label{eq:sums}
\end{align}
where we have applied Kolmogorov's forward equation~\eqref{eq:forward} to calculate $p'_{x,y}(s)$ and backward equation~\eqref{eq:backward} to calculate $p'_{y+1,z}(h-s)$. The first summation in ~\eqref{eq:sums} can be simplified as
\begin{align*}
    \text{first summation  in ~\eqref{eq:sums}} &= \sum_{y=0}^x\sum_{z=j}^{y+1} q_{y+1,y}p_{x,y+1}(s)p_{y+1,z}(h-s)- \sum_{y=0}^x\sum_{z=j}^{y+1} q_{y+1,y}p_{x,y}(s)p_{y,z}(h-s)\\
 &= \sum_{y=1}^{x+1}\sum_{z=j}^{y} q_{y,y-1}p_{x,y}(s)p_{y,z}(h-s)- \sum_{y=0}^x\sum_{z=j}^{y+1} q_{y+1,y}p_{x,y}(s)p_{y,z}(h-s)\\
&= \sum_{y=1}^{x}\sum_{z=j}^{y} [q_{y,y-1}-q_{y+1,y}]p_{x,y}(s)p_{y,z}(h-s)\\
&~~~~~~~~+\sum_{z=j}^{x+1} q_{x+1,x}p_{x,x+1}(s)p_{x+1,z}(h-s)-\sum_{z=j}^{1} q_{1,0}p_{x,0}(s)p_{0,z}(h-s)\\
&= \sum_{y=1}^{x}\sum_{z=j}^{y} [q_{y,y-1}-q_{y+1,y}]p_{x,y}(s)p_{y,z}(h-s),
\end{align*}
where for the last equality we use $p_{x,x+1}(s)=p_{0,1}(h-s)=0$ according to the definition of the pure departure process. Therefore
\begin{align*}
F'(s) & =  \sum_{y=1}^{x}\sum_{z=j}^{y} [q_{y,y-1}-q_{y+1,y}]p_{x,y}(s)p_{y,z}(h-s)+\sum_{y=0}^x\sum_{z=j}^{y+1}[q_{y+1,y}-q_{y,y-1}]p_{x,y}(s)p_{y+1,z}(h-s)\\
&\geq\sum_{y=1}^x\sum_{z=j}^{y}[q_{y+1,y}-q_{y,y-1}]p_{x,y}(s)[p_{y+1,z}(h-s)-p_{y,z}(h-s)]\\
&\geq 0,
\end{align*}

\end{enumerate}
where the first equality is due to the fact
$$q_{y+1,y}-q_{y,y-1}=\min\{y+1,m\}\mu-\min\{y,m\}\mu\geq0,$$
and the second equality is due to part (i) of  Lemma~\ref{lem:pure-departure-2}.

We have proved that $F(s)$ is an increasing function of $s$. Therefore $F(0)\leq F(s)\leq F(h)$ for any $0\leq s\leq h$. Notice that
\begin{align*}
F(0)&=P(\mathcal G_{h}(x+1)\geq j), \quad F(h)=P(\mathcal G_h(x)+1\geq j),
\end{align*}
setting $h=t+s$, and we have
\[
P(\mathcal G_{t+s}(x+1)\geq j)\leq P(\mathcal G_{t}(\mathcal G_s(x)+1)\geq j)\leq P(\mathcal G_h(x)+1\geq j),\quad j\geq1.
\]
\Halmos
\endproof

\section{Proof of Lemma~\ref{lem:pure-departure-3}}\label{proof:pure-departure-3}
\proof{Proof of Lemma~\ref{lem:pure-departure-3}.}
This proof is essentially based on the proof of Lemma~\ref{lem:pure-departure-1}.
(i)  If $x\leq m$, then following the proof of Lemma~\ref{lem:pure-departure-1} (i), initially the $x$ customers are at service at the beginning of the time period. Define Bernoulli random variables $X_i, i=1,\cdots,x$, each for one of the $x$ customers, such that $X_i=1$ if the $i$th customer completes service before time $h$, and $X_i=0$ otherwise. Then $\{X_i, i=1,\cdots,x\}$ are independent, and $P(X_i=1)=1-e^{-\mu h}$. Suppose we pick the server serving the $j$th customer. Then $\sum_{\substack{i=1\\i\neq j}}^x X_i$ is of binomial distribution $\mathrm{Binomial}(x-1,p)$ and independent with $X_j$. Therefore
\begin{align*}
P(\mathcal E|D=d)&=P\left(X_j = 0 \Big| \sum_{i=1}^x X_i = d\right)\\
&= \frac{P\left(X_j=0, \sum_{i=1}^x X_i=d\right)}{P\left(\sum_{i=1}^x X_i=d\right)}\\
&= \frac{P\left(X_j=0\right)P\left( \sum_{\substack{i=1\\i\neq j}}^x X_i=d\right)}{P\left(\sum_{i=1}^x X_i=d\right)}\\
&= \frac{(1-p)\cdot\binom{x-1}{d} p^d(1-p)^{x-1-d}}{\binom{x}{d}p^d(1-p)^{x-d}}\\
&=\frac{x-d}{x}.
\end{align*}
(ii) If $x\geq m+1$, then following the proof of Lemma~\ref{lem:pure-departure-1} (ii),  initially $m$ customers are at service and $x-m$ customers are waiting in the queue. \\
(ii.a) If $0\leq d<x-m$, then $T>h$ according to  Lemma~\ref{lem:pure-departure-1}. In another word, all servers are busy from time 0 to time $h$. If we use $N_i$ to denote the number of departures from the $i$th server, then  $\{N_i, i=1,\cdots,m\}$ are independent Poisson random variables with rate $\lambda h$.  Suppose we pick the server serving the $j$th customer.
 Then $\sum_{\substack{i=1\\i\neq j}}^d N_i$ is of Poisson distribution with rate $(m-1)\lambda$ and independent with $N_j$. Therefore
\begin{align*}
P(\mathcal E|D=d)&=P\left(N_j = 0 \Big| \sum_{i=1}^m N_i = d\right) \\
&= \frac{P\left(N_j = 0, \sum_{i=1}^m N_i = d\right)}{P\left(\sum_{i=1}^m N_i = d\right)}\\
&= \frac{P\left(N_j = 0\right)P\left(\sum_{\substack{i=1\\i\neq j}}^m N_i = d\right)}{P\left(\sum_{i=1}^m N_i = d\right)}\\
&= \frac{e^{-\lambda}\cdot e^{-(m-1)\lambda}[(m-1)\lambda]^d/d!}{e^{-m\lambda}(m\lambda)^d/d!}\\
&= \left(\frac{m-1}{m}\right)^d.
\end{align*}
(ii.b) If $x-m\leq d\leq x$, then $T\leq h$. In another word, all servers are busy from time 0 to $T$ (during this period there are $x-m$ departures), and then the system behaves in the same way as in case (i) (during this period there are $d-x+m$ departures). As event $\mathcal E$ happens if and only if this customer does not complete service during both $(0,T]$ and $(T,h]$, therefore, by combining the argument in (i) and (ii.a), we have
\begin{align*}
P(\mathcal E|D=d)&=\left(\frac{m-1}{m}\right)^{x-m}\times \frac{m-(d-x+m)}{m}\\
&=\left(\frac{m-1}{m}\right)^{x-m}\frac{x-d}{m}.
\end{align*}
\Halmos\endproof

\section{Proof of Theorem~\ref{thm:error}}\label{proof:error}
\proof{Proof of Theorem~\ref{thm:error}}
The proof is built on three arguments presented in Section~\ref{sec:error_single}.

Firstly, we prove that $N_\tau^f\preceq  N_\tau\preceq  N_\tau^b$ for any $\tau\geq0$.

We establish the argument by induction on $\tau$. When $\tau=0$, it holds trivially since $N_0^f=N_0=N_0^b=0$. Assume we have established $N_{\tau-1}^f\preceq  N_{\tau-1}\preceq  N_{\tau-1}^b$ for a given $\tau\geq1$. From time $(\tau-1)h$ to time $\tau h$ there are $A_\tau$ arrivals. Denote the arrival times by $T_1, T_2, \cdots, T_{A_\tau}$ with $(\tau-1)h<T_1<\cdots<T_{A_\tau}<\tau h$. Conditioning on $N_{\tau-1}$, $A_\tau$ and $T_1, T_2, \cdots, T_{A_\tau}$, we observe that the conditional distribution of $N_\tau$ is
\begin{align*}
&N_\tau | \{N_{\tau-1};A_\tau; T_1, T_2, \cdots, T_{A_\tau}\}\\
&\eqd  \mathcal G_{\tau h-T_{A_\tau}}\left(1+\mathcal G_{T_{A_\tau}-T_{A_\tau-1}}\left(\cdots1+\mathcal G_{T_2-T_1}\left(1+ \mathcal G_{T_1-(\tau-1) h} (N_\tau)\right)\cdots\right)\right)\\
&\preceq \mathcal G_{\tau h-T_{A_\tau-1}}\left(1+\mathcal G_{T_{A_\tau-1}-T_{A_\tau-2}}\left(\cdots1+\mathcal G_{T_2-T_1}\left(1+ \mathcal G_{T_1-(\tau-1) h} (N_\tau)\right)\cdots\right)\right)+1\\
&\preceq \cdots\\
&\preceq \mathcal G_{\tau h-T_1}\left(1+ \mathcal G_{T_1-(\tau-1) h} (N_\tau)\right)+A_\tau-1\\
&\preceq \mathcal G_{h}\left(N_{\tau-1}\right) + A_\tau,
\end{align*}
where we use Lemma~\ref{lem:pure-departure-2} (i) and (ii) repeatedly. Furthermore, by assumption, $N_{\tau-1}\preceq  N_{\tau-1}^b$, and using 
Lemma~\ref{lem:pure-departure-2} (i), we have
\begin{align*}
    N_\tau | \{N_{\tau-1};A_\tau; T_1, T_2, \cdots, T_{A_\tau}\}
&\preceq \mathcal G_{h}\left(N_{\tau-1}^b\right) + A_\tau \eqd  N_\tau^b | \{N_{\tau-1}^b;A_\tau;T_1, T_2, \cdots, T_{A_\tau}\}.
\end{align*}
Therefore, we obtain $N_{\tau}\preceq  N_{\tau}^b$ by integrating the conditional probabilities. Similarly, we can also obtain $N_{\tau}^f\preceq  N_{\tau}$. Letting $\tau$ goes to infinity, we can obtain the stochastic dominance relationship in steady state.

Secondly, we prove that $(N_{\tau}^b,D_{\tau+1}^b)\eqd(N_{\tau-1}^f +A_{\tau}, D_{\tau}^f)$ for any $\tau\geq 1$.

Again, we establish the argument by induction on $\tau$. When $\tau=1$, since $D_0^b=N_0^b=0$, we have 
$$N_1^b=N_0^b+A_1-D_1^b=A_1=N_0^f+A_1.$$
In addition, according to the sampling rule of the approximation schemes described in Section~\ref{subsubsec:sim_N},
\begin{align*}
 D_2^b&\sim \mathrm{GenerateDeparture}(N_{1}^b, m, \mu, h),\\
 D_1^f&\sim \mathrm{GenerateDeparture}(N_0^f+A_1, m, \mu, h).
\end{align*}
Given the value of  $N_1^b$ and $N_0^f+A_1$ are the same, the conditional distributions of $D_2^b$ and $D_1^f$ are also the same. 
Therefore, the joint distribution of $(N_1^b, D_2^b)$  and $ (N_0^f +A_1, D_1^f)$ are identical.
Next assume we have established that $(N_{\tau-1}^b, D_\tau^b)$  and $ (N_{\tau-2}^f +A_{\tau-1}, D_{\tau-1}^f)$ are identical in distribution for a given $\tau$, then $N_{\tau-1}^b-D_\tau^b \eqd N_{\tau-2}^f +A_{\tau-1}- D_{\tau-1}^f$. Furthermore, notice that all random variables in this identity only depend on the arrivals up to time $(\tau-1)h$ and hence are independent with $A_\tau$, therefore if we add $A_\tau$ on both sides, the identity still holds. Combining with the updating equations for backward and forward schemes, we have
\begin{align*}
    N_\tau^b &= N_{\tau-1}^b-D_\tau^b+A_\tau\\
    &\eqd N_{\tau-2}^f +A_{\tau-1} - D_{\tau-1}^f+A_\tau\\
    &=N_{\tau-1}^f+A_{\tau}.
\end{align*}
Then according to the sampling rule in Section~\ref{subsubsec:sim_N}, we have $(N_{\tau}^b,D_{\tau+1}^b) \eqd (N_{\tau-1}^f +A_{\tau}, D_{\tau}^f)$.

Thirdly, we prove $E[N_\tau^b]-E[N_\tau^f] =E[D_\tau^f]$, for any $\tau\geq0$.

Since $N_\tau^b$ are $N_{\tau-1}^f + A_\tau$ identical in distribution, we have
$$E[N_\tau^b]-E[N_\tau^f]=E[N_{\tau-1}^f + A_\tau]-E[N_\tau^f]=E[N_{\tau-1}^f + A_\tau-N_\tau^f]=E[D_\tau^f].$$
Since $D_\tau^f$ is the number of customers who finish service and leave the station during the $\tau$th time interval according to the forward approximation, it is bounded from above by the maximum service capacity of the station in this time interval. As there are $m$ exponential servers with service rate $\mu$, the maximum number of customers that can be served with a time length of $h$ is a Poisson random variable with mean $m\mu h$. Therefore
$$E[N_\tau^b]-E[N_\tau^f] = E[D_\tau^f]\leq m\mu h.$$ 
Moreover, in steady state, the average number of departures from the station equals the average number of arrivals to the station, so we have 
$$E[N^b]-E[N^f] = E[D^f]=E[A] = \lambda h.$$
\Halmos
\endproof

\section{Proof of Theorem~\ref{thm:error_network}}\label{proof:error_network}
\proof{Proof of Theorem~\ref{thm:error_network}}
(i) We prove part (i) by induction on the sequence of layers. For any node $i\in\mathcal L^{-1}(1)$, since it is on the first layer and all arriving customers come from outside, the number of customers in this node behaves exactly like a single-station queue. Therefore, we can directly use the results from Theorem~\ref{thm:error}
$$N^f(i)\preceq N(i) \preceq N^b(i), \quad E[N^b(i)]-E[N^f(i)]=\lambda_i h=\tilde \lambda_i h.$$
In addition, from the proof of Theorem~\ref{thm:error} we also known that $D^b(i) \eqd D^f(i)$, where  $D^b(i)$ and $D^f(i)$ are the steady-state number of departures within a time interval for backward and forward schemes, respectively. For both schemes, the departures from node $i$ will go to the nodes in the second layer according to the identical routing matrix $\mathbf P$, therefore the internal arrivals to the second-layer nodes have the identical distribution in both schemes. Now consider a node $i\in\mathcal L^{-1}(2)$. We have shown that $A^{b,in}(i)\eqd A^{f,in}(i)$. Because the external arrivals are independent with the internal arrivals, we have $A^{b,in}(i)+A^{ex}(i)\eqd A^{f,in}(i)+A^{ex}(i)$. Therefore, the steady-state behavior of the number of customers in this node is the same as a single-station queue be interpreting the arrival rate as the total arrival rate (the sum of internal and external arrival rates). Again, we use the results from Theorem~\ref{thm:error} and get
$$N^f(i)\preceq N(i) \preceq N^b(i), \quad E[N^b(i)]-E[N^f(i)]=\tilde \lambda_i h.$$
By induction on the sequence of layers, we can establish the results for all node $i$.

(ii) From Assumption~\ref{asm:network}, we know the queueing network is an open Jackson network. Therefore $\{N(i), i=1,\ldots,n\}$ are independent, so be their Euler approximations $\{N^b(i), i=1,\cdots,n\}$ and $\{N^f(i), i=1,\cdots,n\}$.
As $N=\sum_{i=1}^n N(i)$, $N^b=\sum_{i=1}^n N^b(i)$, $N^f=\sum_{i=1}^n N^f(i)$, from part (i) we get $N^f\preceq N \preceq N^b$. Furthermore,
\begin{align*}
E[N^b]-E[N^f]&= \sum_{i=1}^n \left(E[N^b(i)] -E[N^f(i)]\right)=\sum_{k=1}^l \sum_{i\in\mathcal L_k} \tilde \lambda_i.
\end{align*}
Since $\sum_{i\in\mathcal L_1} \tilde \lambda_i  = \Lambda_1$, and for $k\geq 2$ we have
\begin{align*}
\sum_{i\in\mathcal L_k} \tilde \lambda_i 
&= \sum_{i\in\mathcal L_k} ( \lambda_i+\sum_{j\in \mathcal L_{k-1}} \tilde \lambda_j p_{ji})= \Lambda_k+\sum_{j\in \mathcal L_{k-1}} \tilde \lambda_j \sum_{i\in\mathcal L_k} p_{ji}\leq \Lambda_k + \sum_{j\in \mathcal L_{k-1}} \tilde \lambda_j,
\end{align*}
by induction on $k$ we obtain
\[
\sum_{i\in\mathcal L_k} \tilde \lambda_i \leq \sum_{i=1}^k \Lambda_i, \quad k=1,\cdots,l.
\]
Therefore
\begin{align*}
E[N^b]-E[N^f]&\leq \sum_{k=1}^l \sum_{i=1}^k \Lambda_i = \sum_{i=1}^k (l+1-k) \Lambda_i.
\end{align*}
\Halmos
\endproof

\section{Proof of Theorem~\ref{thm:relative_error}}\label{proof:relative_error}
\proof{Proof of Theorem~\ref{thm:relative_error}.} According to Theorem~\ref{thm:error}, the absolute errors are no more than $\lambda h$. On the other hand, it is known that the steady-state average number of customers in the system can be expressed as 
\[{E[N]} = m\rho  + \frac{\rho }{{1 - \rho }}C\left( {m,\rho } \right).\] 
Here $C\left( {m,\rho} \right)$ is known as the Erlang C formula, which is the probability that an arriving customer finds all servers busy. Therefore
\[
\overline{\mathrm{RE}}(h)= \frac{\lambda h}{E[N]}=\frac{\rho m\mu h}{m\rho+\frac{\rho}{1-\rho} C(m,\rho)}= \frac{\mu h}{1+\frac{C(m,\rho)}{m(1-\rho)}}.
\]
As $C(m,\rho)\geq 0$, we have $\overline{\mathrm{RE}}(h)\leq \mu h$. Furthermore, consider different asymptotic regimes.
\begin{itemize}
\item If $m(1-\rho)\rightarrow \infty$, as $C(m,\rho)\in[0,1]$, we have 
$$\frac{\overline{\mathrm{RE}}(h)}{\mu h} = \frac{1}{1+\frac{C(m,\rho)}{m(1-\rho)}}\rightarrow 1.$$ 

\item If $m(1-\rho)\rightarrow \beta\in(0,\infty)$, as $m\rightarrow\infty$, we mush have $\sqrt m(1-\rho)\rightarrow 0$. In this case, according to \cite{halfinwhitt1981}, we have $C(m,\rho)\rightarrow 1$. Therefore
\[
\frac{\overline{\mathrm{RE}}(h)}{\mu h}\rightarrow \frac{1}{1+1/\beta}=\frac{\beta}{1+\beta}.
\]

\item If $m(1-\rho)\rightarrow 0$, we also have $\sqrt m(1-\rho)\rightarrow 0$, and hence $C(m,\rho)\rightarrow 1$. Therefore
$$\frac{\overline{\mathrm{RE}}(h)}{m(1-\rho)\mu h}= \frac{1}{m(1-\rho)+C(m,\rho)} \rightarrow 1.$$
\end{itemize}
\Halmos
\endproof

\section{Proof of Theorem~\ref{thm:relative_error_network}}\label{proof:relative_error_network}
\proof{Proof of Theorem~\ref{thm:relative_error_network}.}
Without loss of generality, consider the backward scheme, and the relative error
\begin{align*}
    \frac{{E\left[ N_\tau^{b} \right] - E\left[ N_\tau \right]}}{{E [N_\tau]}}
    &=\frac{{\sum_{i=1}^n \left(E[ N_{\tau,i}^{b}] -  E\left[ N_{\tau,i} \right]\right)}}{\sum_{i=1}^n{E [N_{\tau,i}]}}\\
    &\leq \frac{{\sum_{i=1}^n \left(E[ N_{\tau,i}^{b}] -  E[ N_{\tau,i}^f ]\right)}}{\sum_{i=1}^n{E [N_{\tau,i}]}}\\
    &= \frac{{\sum_{i=1}^n \tilde \lambda_i h}}{\sum_{i=1}^n{E [N_{\tau,i}]}} = \overline{\mathrm{RE}}(h).
\end{align*}
In addition,
\begin{align*}
    \overline{\mathrm{RE}}(h)&= \frac{{\sum_{i=1}^n m_i\rho_i\mu_i h}}{\sum_{i=1}^n{\left[m_i\rho_i+\frac{\rho_i}{1-\rho_i}C(m_i,\rho_i)\right]}}\leq \frac{{\sum_{i=1}^n m_i\rho_i\mu_i }}{\sum_{i=1}^n{m_i\rho_i}}h.
\end{align*}
Notice that the term before $h$ is a weighted average of $\mu_i$, therefore it is no more than
$\left(\max_{1\leq i\leq n}\mu_i\right)$.
\Halmos
\endproof

\end{APPENDICES}

\end{document}